\documentclass[11pt,a4paper]{article} 

\usepackage{titlesec}
\usepackage{color}

\usepackage[utf8]{inputenc}
\usepackage[english]{babel}
\usepackage[T1]{fontenc} 
\usepackage{amsfonts} 

\usepackage{graphicx}
\graphicspath{{Images/}}
\usepackage{eso-pic} 
\usepackage{caption} 
\usepackage{transparent}

\usepackage{amsmath}
\usepackage{amsthm}
\usepackage{bm}
\usepackage[overload]{empheq}  

\usepackage{tabularx}
\usepackage{longtable} 
\usepackage{colortbl}

\usepackage{algorithm}
\usepackage{algorithmic}

\usepackage[colorlinks=true,linkcolor=black,anchorcolor=black,citecolor=black,filecolor=black,menucolor=black,runcolor=black,urlcolor=black]{hyperref} 
\usepackage{cleveref}
\usepackage[square, numbers, sort&compress]{natbib} 
\usepackage{appendix}

\usepackage{enumitem}

\usepackage{amsthm,thmtools,xcolor} 
\usepackage{comment} 
\usepackage{fancyhdr} 
\usepackage{lipsum} 
\usepackage{tcolorbox} 

\usepackage{svg}
\usepackage{subcaption}

\newcommand{\bea}{\begin{eqnarray}} 
\newcommand{\eea}{\end{eqnarray}}

\usepackage[bottom=2.0cm,top=2.0cm,left=2.0cm,right=2.0cm]{geometry}
\usepackage{lmodern}

\definecolor{bluePoli}{cmyk}{0.4,0.1,0,0.4}

\declaretheoremstyle[
  headfont=\color{bluePoli}\normalfont\bfseries,
  bodyfont=\color{black}\normalfont\itshape,
]{colored}

\theoremstyle{colored}

\newcounter{algsubstate}

\newcolumntype{L}[1]{>{\raggedright\let\newline\\\arraybackslash\hspace{0pt}}m{#1}}
\newcolumntype{C}[1]{>{\centering\let\newline\\\arraybackslash\hspace{0pt}}m{#1}}
\newcolumntype{R}[1]{>{\raggedleft\let\newline\\\arraybackslash\hspace{0pt}}m{#1}}

\setlist[itemize,1]{label=$\bullet$}
\setlist[itemize,2]{label=$\circ$}
\setlist[itemize,3]{label=$-$}
\setlist{nosep}

\newcommand\BackgroundPic{
	\put(237,365){
	    \parbox[b][\paperheight]{\paperwidth}{%
	    \vfill
		\centering
		\transparent{0.4}
		\includegraphics[width=0.44\paperwidth]{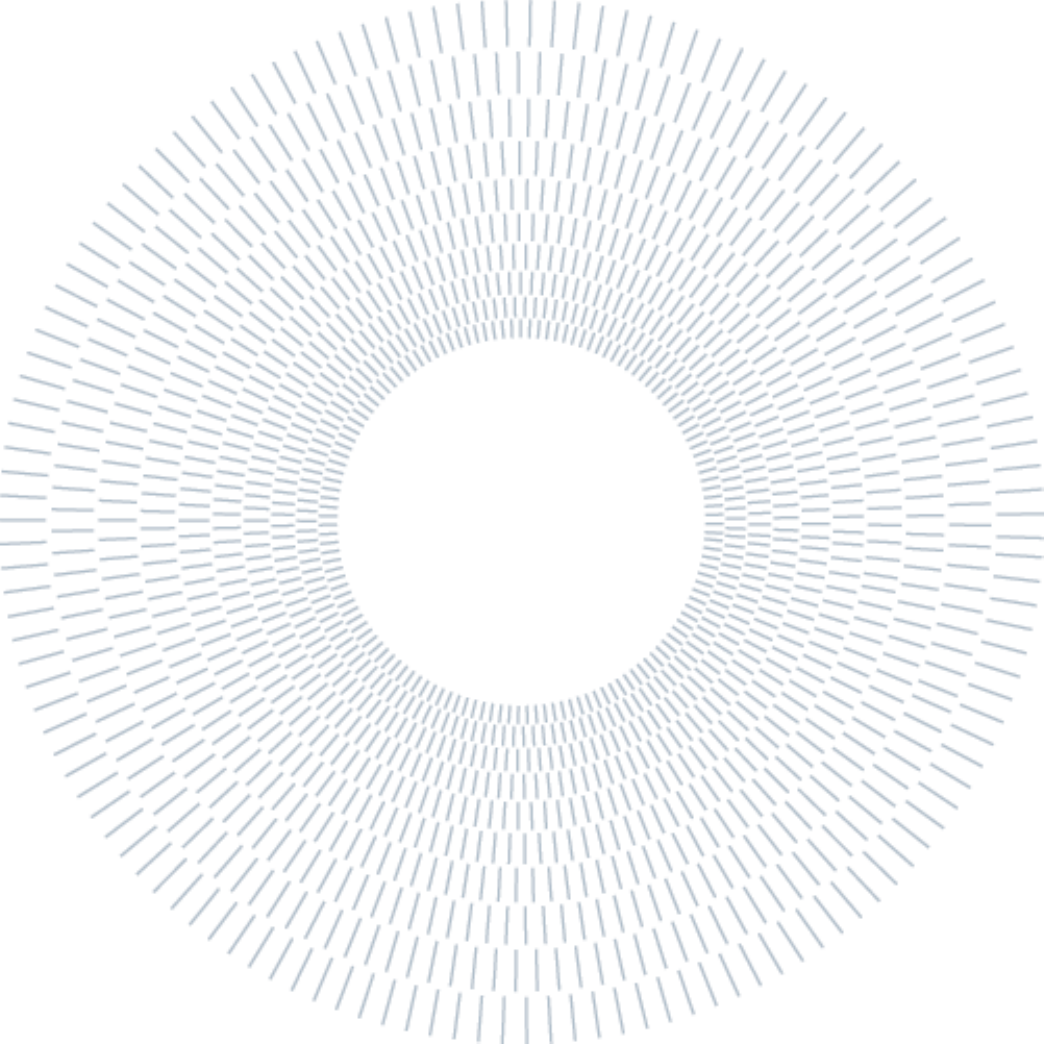}%
		\vfill}
		}
}

\titleformat{\section}
{\color{bluePoli}\normalfont\Large\bfseries}
{\color{bluePoli}\thesection.}{1em}{}
\titlespacing*{\section}
{0pt}{3.3ex}{3.3ex}

\titleformat{\subsection}
{\color{bluePoli}\normalfont\large\bfseries}
{\color{bluePoli}\thesubsection.}{1em}{}
\titlespacing*{\subsection}
{0pt}{3.3ex}{3.3ex}

\makeatletter
\patchcmd{\headrule}{\hrule}{\color{black}\hrule}{}{} 
\patchcmd{\footrule}{\hrule}{\color{black}\hrule}{}{} 
\makeatother

\newcommand{\figref}[1]{(Fig.~\ref{#1})}

\renewcommand{\title}{Algorithms and optimizations for global non-linear hybrid
fluid-kinetic finite element stellarator simulations}
\renewcommand{\author}{Luca Venerando Greco}
\newcommand{\course}{High Performance Computing Engineering - Ingegneria del Supercalcolo}
\newcommand{\advisor}{Prof. Paola F. Antonietti}
\newcommand{\firstcoadvisor}{Dr. Matthias Hoelzl} 
\newcommand{\secondcoadvisor}{Prof. Pierre Talbot} 
\newcommand{\ID}{245265}
\newcommand{\YEAR}{2024-2025}
\renewcommand{\abstract}{Predictive modeling of stellarator plasmas is crucial for advancing nuclear fusion energy, yet it faces unique computational difficulties. One of the main challenges is accurately simulating the dynamics of specific particle species that are not well captured by fluid models, which necessitates the use of hybrid fluid-kinetic models. The non-axisymmetric geometry of stellarators fundamentally couples the toroidal Fourier modes, in contrast to what happens in tokamaks, requiring different numerical and computational treatment.

This work presents a novel, globally coupled projection scheme inside the JOREK finite element framework. The approach ensures a self-consistent and physically accurate transfer of kinetic markers to the fluid grid, effectively handling the complex 3D mesh by constructing and solving a unified linear system that encompasses all toroidal harmonics simultaneously. To manage the computational complexity of this coupling, the construction of the system's matrix is significantly accelerated using the Fast Fourier Transform (FFT). The efficient localization of millions of particles is made possible by implementing a 3D R-Tree spatial index, which supports this projection and ensures computational tractability at scale.

On realistic Wendelstein 7-X stellarator geometries, the fidelity of the framework is rigorously shown. In sharp contrast to the uncoupled approaches' poor performance, quantitative convergence tests verify that the coupled scheme attains the theoretically anticipated spectral convergence. 

This study offers a crucial capability for the predictive analysis and optimization of next-generation stellarator designs by developing a validated, high-fidelity computational tool.}

\newcommand{\keywords}{stellarator Simulation, Hybrid Fluid-Kinetic Model, Finite Element Method (FEM), Particle-in-Cell (PiC), JOREK, High Performance Computing (HPC)}

\begin{document}


\AddToShipoutPicture*{\BackgroundPic}

\hspace{-0.6cm}\includegraphics[width=0.6\textwidth]{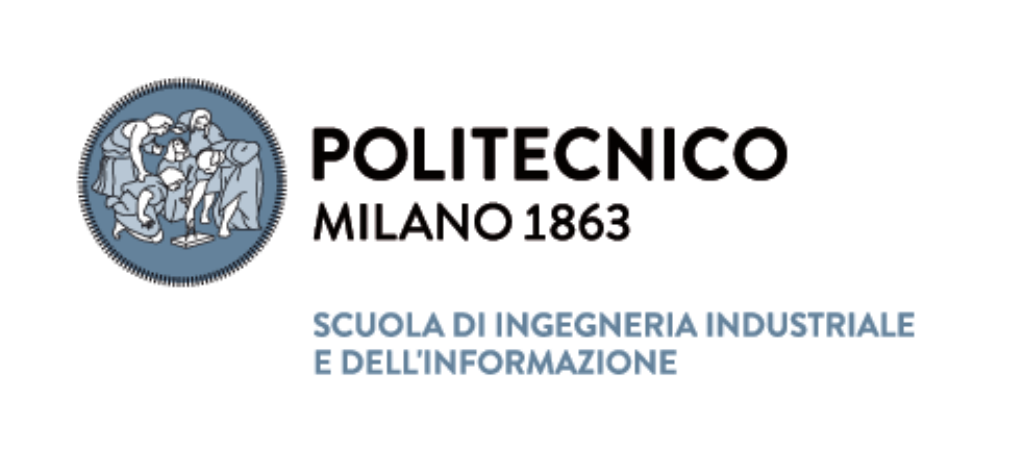}

\vspace{-1mm}
\Large{\textbf{\color{bluePoli}{\title}}}\\

\vspace{-0.2cm}
\fontsize{0.3cm}{0.5cm}\selectfont \bfseries \textsc{\color{bluePoli} Tesi di Laurea Magistrale in \\ \course}\\

\vspace{-0.2cm}
\large{\textbf{\author, \ID}}

\small \normalfont

\vspace{11pt}

\centerline{\rule{1.0\textwidth}{0.4pt}}

\begin{center}
\begin{minipage}[t]{.24\textwidth}
\begin{minipage}{.90\textwidth}
\noindent
\scriptsize{\textbf{Advisor:}} \\
\advisor \\
\\
\textbf{Co-advisors:} \\ 
\firstcoadvisor \\ 
\secondcoadvisor \\ 
\\ 
\textbf{Academic year:} \\
\YEAR \\
\\
\end{minipage}
\end{minipage}
\begin{minipage}{.74\textwidth}
\noindent \textbf{\color{bluePoli} Abstract:} {\abstract}
\end{minipage}
\end{center}

\vspace{15pt}

\begin{tcolorbox}[arc=0pt, boxrule=0pt, colback=bluePoli!60, width=\textwidth, colupper=white]
    \textbf{Key-words:} \keywords
\end{tcolorbox}

\vspace{12pt}


\newpage
\tableofcontents
\newpage
\section{Introduction}
\label{sec:introduction}

Nowadays, nuclear fusion is one of the most important scientific and engineering problems of our time due to the quest for clean, sustainable energy. Fusion reactors promise an almost infinite energy source with major safety and environmental advantages over traditional fission and fossil fuel technologies by exploiting a similar process that powers stars \cite{ITER}. The confinement of a high-temperature plasma in a magnetic field is the most promising method for accomplishing controlled fusion on Earth. However, high-fidelity computational simulations are required due to the enormous expense and complexity of building and running experimental reactors \cite{FSP}. These simulations are essential resources for developing new approaches, deciphering experimental data, and investigating plasma phenomena that are challenging or impossible to measure directly.

Fluid and kinetic models are the two main categories into which plasma simulation paradigms fall. Large-scale instabilities and general plasma behavior are effectively captured by fluid models, which are governed by Magneto-HydroDynamics (MHD) equations and treat the plasma as a continuum \cite{FluidSolvers}. Kinetic models, on the other hand, follow the paths of individual particles and offer a more accurate, albeit more computationally expensive, representation of the plasma \cite{KineticSolvers}. The development of hybrid fluid-kinetic approaches has been prompted by the prohibitive computational cost of fully kinetic simulations for reactor-scale plasmas over MHD-relevant timescales. These models create a self-consistent framework by combining the accuracy of a kinetic treatment for certain, important particle populations with the computational efficiency of a fluid description for the bulk plasma \cite{HybridSolvers}.

Numerous specialized numerical codes have been created to handle the various physical regimes found in magnetically confined plasmas. Usually, each of these tools is tailored for a particular set of scales, equations, or configurations. In particular, for hybrid MHD-kinetic simulations many codes have been developed, for instance: JOREK \cite{JOREK}, NIMROD \cite{nimrod}, M3D-C1-K \cite{M3D-C1, M3D-C1-K}, MEGA \cite{mega} and XHMGC \cite{xhmgc} are some of the leading codes in the field, designed to couple a fluid description of the bulk plasma with a kinetic treatment for specific particle species. Among them, specific interest has sparked in the simulation of stellarator reactors, leading to the development of specific extensions to account for the three-dimensionality of the plasma \cite{JOREK3D, M3D-stell, nimstell}, underscoring the interest and need of the scientific community for this kind of reactor.

The tokamak and the stellarator are the two main device concepts that have emerged in the field of magnetic confinement fusion. Both configurations are actively researched as promising concepts for a fusion reactor with different strengths and weaknesses. The tokamak's intrinsic toroidal symmetry makes its physical model and numerical implementation easier, but the stellarator's non-axisymmetric, three-dimensional magnetic field poses a much greater computational challenge. The need for sophisticated simulation tools that can faithfully capture the unique physics of stellarator designs is growing in importance, triggered in particular by the successful operation of the W7-X. As a result, although there are established hybrid simulation frameworks for tokamaks in codes like JOREK \cite{JOREK}, there is still a vital and ongoing research need to extend them to stellarator geometries.

The breakdown of toroidal axisymmetry is the main obstacle to applying the available hybrid models to stellarators. This symmetry permits a spectral decomposition of the mesh geometry in the toroidal direction, in which Fourier harmonics are decoupled, in tokamak simulations. The numerical problem is greatly simplified by this decoupling since each harmonic's kinetic particle data can be independently projected onto the fluid grid. These harmonics are intrinsically coupled in a stellarator due to the three-dimensional nature of the geometry. The basis functions utilized for the finite element discretization are functions of all three spatial dimensions since the physical form of the poloidal cross-section varies with the toroidal angle. The uncoupled projection scheme would produce nonphysical results if applied to stellarators and would fail to accurately represent the physical plasma.

By creating, putting into practice, and validating the algorithms necessary for a global non-linear hybrid fluid-kinetic simulation capability within the JOREK stellarator model, this thesis directly tackles this challenge. The design of a novel particle-in-cell projection scheme that completely takes into account the coupling between toroidal harmonics that results from the non-axisymmetric mesh geometry is the main contribution. This approach allows for a precise and self-consistent transfer of information from the kinetic particles to the fluid grid by accurately formulating and solving the global linear system for all harmonics at the same time. These projections are needed for diagnosing the particle dynamics and for a self-consistent coupling between fluid and kinetics. To overcome the significant computational cost of this approach, two key optimizations were implemented. Firstly, the assembly of the global system matrix is accelerated using the Fast Fourier Transform (FFT). Secondly, a three-dimensional R-Tree spatial index is introduced to handle the expensive task of localizing a high number of particles within the complex 3D grid.

The work is structured as follows: the required theoretical background is given in section \ref{sec:background}, which also describes the $G^1$ Bézier finite element framework used in JOREK for the fluid model and the reduced-MHD equations. The kinetic particle-in-cell methodology is then presented, along with an original derivation of the mathematical formulation for the projection scheme and the Boris pusher for particle tracing. The main contributions of this study are described in detail in section \ref{Chapter3}. The new coupled-harmonic projection algorithm is implemented after first validating the basic particle tracing in stellarator configurations. Quantitative convergence tests and qualitative projections on realistic stellarator geometries are used to illustrate the precision and effectiveness of this new scheme. Following this validation, the chapter analyzes the computational performance of the key optimizations, presenting a speedup analysis of the FFT-accelerated matrix construction against the direct integration method. We conclude by exploring how the 3D R-Tree for effective particle localization is implemented and examining its performance. This methodical approach paves the way for next-generation, high-fidelity stellarator simulations by validating the new framework as a reliable and accurate tool.

\section{Background} 
\label{sec:background}

The theoretical foundation for comprehending the hybrid simulation approach used in this thesis is laid out in this section. We first describe the motivation behind this study, emphasizing the significance of using the stellarator and tokamak models for high-fidelity plasma simulations. Next, we describe the hybrid model's two main pillars. The reduced MHD equations and the numerical framework for solving them in the JOREK code, which combines $G^1$ continuous Bézier finite elements for the poloidal plane with a toroidal Fourier expansion, are introduced first in the fluid component. Second, we present the kinetic model, which explains how the Newton-Lorentz equations govern the motion of individual particles and how the Boris pusher algorithm is used to integrate them. 

\subsection{Motivation}
Fusion reactors have been a primary focus of clean energy research for the past half-century, offering the promise of clean, virtually limitless energy without the drawbacks associated with fission energy. The significant engineering challenges and extensive resources required for experimentation have slowed progress, necessitating computer-aided simulations to test and validate hypotheses.

\subsubsection{Simulation types}

A basic taxonomy of plasma physics simulations was proposed in early literature \cite{Plasma_physics_via_computer_simulation}, establishing a clear distinction between kinetic and fluid simulations. Kinetic simulations describe plasma at the particle level, while fluid simulations treat plasma as a single or multiple fluids governed by MHD equations. The fundamental trade-off between these approaches balances computational requirements against accuracy. While fully kinetic simulations can precisely capture particle interactions within plasma, the computational complexity grows exponentially with particle count. On the other hand, fluid-based simulations solve Maxwell's equations on finite element grids, with complexity directly related to grid resolution and the sophistication of the MHD equations employed. In recent years, despite the increase in HPC computational power\cite{trends}, the inherent approximations in fluid simulations have limited achievable accuracy, driving the development of hybrid simulation approaches.
 
\begin{figure}
    \centering
    \includegraphics[width=0.75\linewidth]{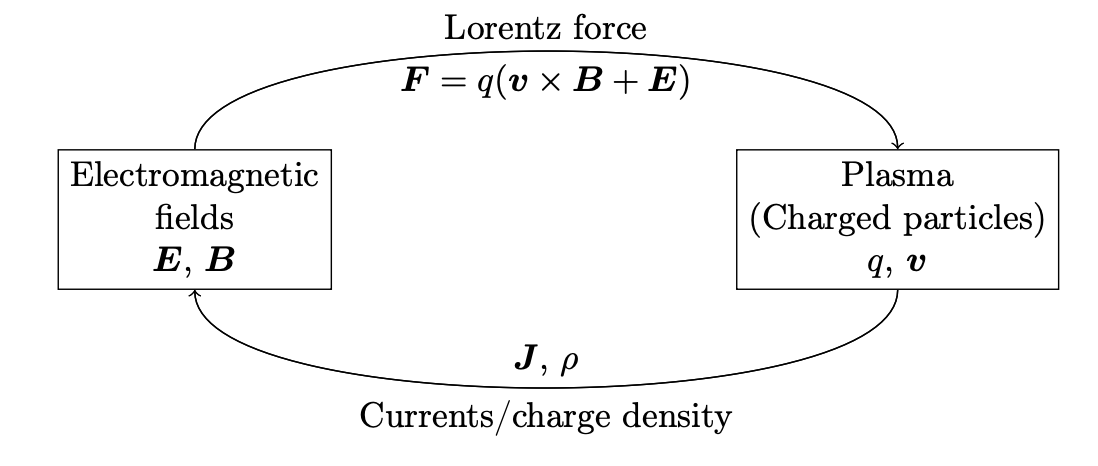}
    \caption{Feedback loop between electromagnetic fields and plasma in kinetic-fluid simulations. The Lorentz force acts on charged particles, influencing their motion and altering currents and charge densities, which in turn affect the electromagnetic fields \cite{Introduction_to_stellarators}.}
    \label{fig:coupling_loop}
\end{figure}

In these hybrid simulations, the fluid solver is combined with a kinetic simulation of particular particle species that cannot be described well in the fluid picture, like supra-thermal ions \cite{Bogaarts}, neutral particles or impurities \cite{sven_impurities}, or runaway electrons \cite{Hannes_RE}. The time evolution of the kinetic simulation provides accurate information about the time evolution of plasma constituents under kinetic assumptions, which is then incorporated into the next fluid time step by coupling terms. A schematic example of the underlying physics is shown in Fig. \ref{fig:coupling_loop}. This creates a continuous feedback loop between the two simulation approaches, allowing them to self-consistently evolve the complete system.

\subsubsection{Reactors types}

Fusion reactions exploiting magnetic confinement of the plasma are carried out at the moment, mainly, using two different types of reactors, stellarator and tokamak. The latter has seen widespread use in the scientific community for its symmetric shape and ease of design; on the other hand, designs and simulations of stellarator reactors have been limited by the computational power available \cite{stellarator_understudied}. Nevertheless, the stellarator reactor offers a number of advantages such as intrinsically steady-state operation, fewer MHD instabilities, and nearly disruption-free operation, i.e., no severe loss of the overall plasma confinement quality \cite{tokamak_vs_stellarator}.

\begin{figure}[h]
    \centering
    \includegraphics[width=0.75\linewidth]{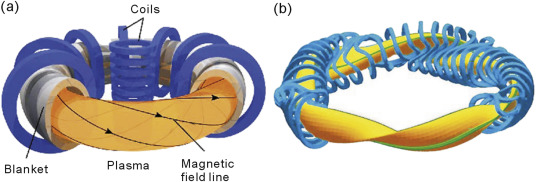}
    \caption{A computer graphic representation of the two different types of reactors \cite{tokamak_vs_stellarator}. On the left, the tokamak; on the right, the stellarator.}
    \label{fig:tok_vs_stel}
\end{figure}

Due to the popularity of tokamak designs, significant progress has been made in developing physically accurate simulations that combine finite element method (FEM) resolution of (reduced) MHD models with kinetic solvers, creating hybrid solutions, e.g., in the JOREK \cite{JOREK} codebase that is of interest to this thesis. While substantial work has been done to develop MHD models for stellarator reactors \cite{Nikita1, Nikita2}, current simulations lack the crucial kinetic solver component. Integrating such a solver with fluid models would enable more accurate stellarator simulations, potentially accelerating research for the next generation of these less widely studied reactor designs.

\subsection{Governing equations and Numerical Approximations for the fluid}

The code is logically separated into two different parts: the fluid part and the kinetic part.

The fluid part provides a starting point for the kinetic module by solving a reduced MHD model of the plasma using the Finite Element Method (FEM). 

\subsubsection{The MHD stellarator model}

Firstly conceptualized in \cite{Nikita1} and further implemented in \cite{Nikita2}, the implemented model aims at handling the MHD equations by not considering the fast magnetosonic waves, since they account for the fastest time scales of the system and are often not relevant to the problems considered.

The set of equations derived from the viscoresistive MHD equations is provided with time evolution equations for the density, velocity, magnetic field, and internal energy, as well as additional definitions and constraints, as follows:

\begin{equation}
\begin{gathered}
\displaystyle \frac{\partial \rho}{\partial t} + \nabla \cdot (\rho \mathbf{v}) = P, \\[3ex]
\displaystyle \rho \frac{\partial \mathbf{v}}{\partial t} + \rho(\mathbf{v} \cdot \nabla)\mathbf{v} + \mathbf{v}P = \mathbf{j} \times \mathbf{B} - \nabla p + \mu \Delta \mathbf{v},\\[3ex] \frac{\partial \mathbf{B}}{\partial t} = -\nabla \times \mathbf{E}, \\[3ex]
\displaystyle \frac{\partial \mathcal{E}}{\partial t} + \nabla \cdot \left[ \left( \frac{\rho v^2}{2} + \frac{\gamma p}{\gamma - 1} \right) \mathbf{v} + \frac{p}{\gamma-1} \frac{D_\perp}{\rho} \nabla_\perp \rho + \frac{\mathbf{E} \times \mathbf{B}}{\mu_0} - \kappa_\perp \nabla_\perp T - \kappa_\parallel \nabla_\parallel T \right] = S_e - \frac{v^2 P}{2}, \\[3ex]
\displaystyle \mathcal{E} = \frac{\rho v^2}{2} + \frac{p}{\gamma-1} + \frac{B^2}{2\mu_0},\\[3ex]
\nabla \times \mathbf{B} = \mu_0 \mathbf{j}, \\[3ex]
\nabla \cdot \mathbf{B} = 0,\\[3ex]
\mathbf{E} = -\mathbf{v} \times \mathbf{B} + \eta \mathbf{j}, \\[3ex]
P = \nabla \cdot (D_\perp \nabla_\perp \rho) + S_\rho.
\end{gathered}
\end{equation}

These equations, modified with suitable ansatzes, will be solved on a 2D $G^1$ continuous finite element grid combined with a toroidal Fourier decomposition to provide the electromagnetic field in which the kinetic particles will move through.
For a definition of all symbols, refer to \cite{Nikita1}.

\subsubsection{Spatial Discretization}

The JOREK code uses $G^1$ continuous isoparametric Bézier finite elements \cite{JOREK} to discretize the poloidal cross-section of the plasma and a Fourier expansion on the toroidal direction. This allows for achieving the same physical accuracy with a low number of DoFs compared to other types of elements \cite{BezierFEM}.

\subsubsection{Bézier Finite Elements}

For the finite element method taken into account, Bernstein polynomials are used along each of the two element-local coordinates $s$ and $t$:

\begin{equation}
B^{3}_{i}(s) = \frac{3!}{i!\,(3 - i)!} \, s^{i}\,(1 - s)^{3 - i}
\end{equation}

These exhibit some interesting properties such as:
\begin{itemize}
    \item $\left(B^n_i\right)$ is a basis of $\mathbb{P}^n$, 
    \item $0 \leq B^n_i(s) \leq 1, \forall s \in \left[0,1\right]$
    \item $\left(B^n_i\right)$ is a partition of unity $\implies$ $\sum^{n}_{i=0}B^n_i(s) = 1, \forall s \in \left[0,1\right]$
\end{itemize}

\vspace{5pt}

Each element uses two different third-degree Bernstein polynomials defined by:

\begin{equation}
B^{3}_{i,j}(s,t) = B^{3}_{i}(s)\, B^{3}_{j}(t) \qquad i,j = 0 \ldots 3,
\end{equation}

Where $0 \leq s, t \leq 1$ denote the local coordinates, which take values 0 and 1 at the four element vertices

These 16 basis functions $B^{3}_{i,j}(s,t)$ are then used to build a 2D Bézier patch through parametrization of 16 control points $P_{i,j}$, each of them belonging to an appropriate vector space, in our case $P_{i,j} \in \mathbb{R}$.

The position of each point $P(s,t)$ on the patch is defined as:

\begin{equation}
P(s,t) = \sum_{i=0}^{3} \sum_{j=0}^{3} P_{i,j} B^{3}_{i,j}(s,t) \label{eq:Bézier_def}
\end{equation}

In this equation, the Bernstein polynomial serves as the weights for the control points. Since the used polynomials form a partition of unity, some properties arise, in particular:

A point $P(s,t)$ on a parametric surface is a convex combination of the set of control points $\{P_{i,j}\}$ if it can be written as:

\begin{equation}
P(s,t) = \sum_{i=0}^{3} \sum_{j=0}^{3} \alpha_{i,j}(s,t) P_{i,j}
\end{equation}

Where the blending functions $\alpha_{i,j}(s,t)$ must satisfy the following properties for all $(s,t)$ in their domain (e.g., $s,t \in [0,1]$):
\begin{itemize}
    \item \textbf{Non-negativity:} $\alpha_{i,j}(s,t) \ge 0 \quad \forall i,j$\vspace{4pt}
    \item \textbf{Partition of Unity:} $\sum_{i=0}^{3} \sum_{j=0}^{3} \alpha_{i,j}(s,t) = 1$ \vspace{4pt}
\end{itemize}

Since the polynomials satisfy both conditions, it is guaranteed that all the points are contained in the convex hull of their control points.

From \eqref{eq:Bézier_def} we can also derive the local derivatives, for instance on $P_{0,0}$:

\begin{align}
\frac{\partial P}{\partial s}\bigg|_{(s,t)=(0,0)} &= 3(P_{1,0} - P_{0,0}); \\
\frac{\partial P}{\partial t}\bigg|_{(s,t)=(0,0)} &= 3(P_{0,1} - P_{0,0}); \\
\frac{\partial^2 P}{\partial s \partial t}\bigg|_{(s,t)=(0,0)} &= 9(P_{1,1} + P_{0,0} - P_{0,1} - P_{1,0}). 
\label{eq:derivates_in_00}
\end{align}

\begin{figure}
    \centering
    \includegraphics[width=0.5\linewidth]{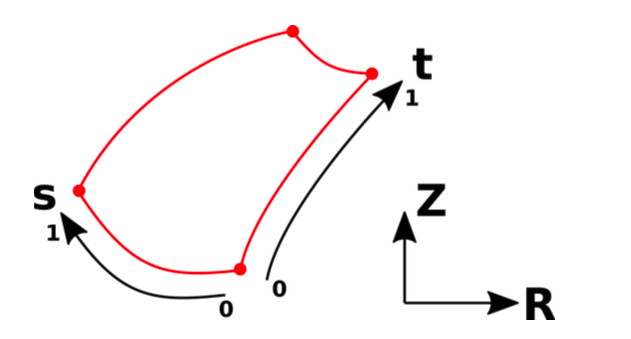}
    \caption{A single 2D isoparametric Bézier element in an axisymmetric R-Z space. The geometry is parameterized by a mapping from the local s-t parent domain using Bernstein polynomial basis functions, which are also used to interpolate field variables like temperature.}
    \label{fig:element_local}
\end{figure}

Both variables and space are discretized using these 16 basis functions, giving the iso-parametric characterization to this method.

\subsubsection{Tesselation of the domain}

In the finite element method, we split the entire domain into different smaller elements, in our case Bézier patches (e.g., Fig. \ref{fig:element_local}). In doing so, we need to take care to satisfy specific conditions imposed by the application's domain-specific requirements. 

The first important constraint to satisfy is Geometric continuity between elements ($G^0$), which simply ensures that there are no gaps between two adjacent elements. Without loss of generality, we consider two adjacent surfaces $S$ and $S'$ which share a common boundary, assuming such boundary lies at the edge $s = 1$ for patch $S$, and $s = 0$ for patch $S'$. 

The boundary of the first patch is defined by $\{P_{3,0}, P_{3,1}, P_{3,2}, P_{3,3}\}$, while for the second one $\{P'_{0,0}, P'_{0,1}, P'_{0,2}, P'_{0,3}\}$. $G^0$ is therefore easily achieved by simply enforcing:

\begin{equation}
P_{3,j} = P'_{0,j} \qquad \forall j \in {0, 1, 2, 3}
\end{equation}

Having a continuous surface across the domain is almost trivial, and given for most elements, it becomes more interesting to consider the $G^1$ continuity between the elements. This must be enforced to avoid sharp, non-physical creases between elements. To ensure such smooth surfaces between elements, the derivatives need to be constrained in some way. $G^1$ continuity requires satisfaction of the so-called "tangent plane condition" \cite{tangent_plane}

The tangent plane at any point of the boundary is spanned by the four partial derivatives. For instance, considering the previous shared edge, assuming $G^0$ continuity, we know that:
\begin{equation}
    \frac{\partial P}{\partial t}\bigg|_{s=0} = \frac{\partial P'}{\partial t'}\bigg|_{s'=0}
\end{equation}

Leaving us with the need to constrain only the other derivatives:

\begin{equation}
k_1(t)\left(\frac{\partial P}{\partial s}\right) + k_2(t)\left(\frac{\partial P}{\partial t}\right) + k_3(t)\left(\frac{\partial P'}{\partial s'}\right) = 0.
\end{equation}

In practice, this requirement is equivalent to imposing the collinear alignment of the control points across the boundary \cite{continuity_bezier}:
\begin{equation}
P_{3,j} - P'_{1,j} = k (P_{2,j} - P_{3,j}) \qquad \forall j \in {0, 1, 2, 3} 
\label{eq:g1_continuity}
\end{equation}

In \eqref{eq:g1_continuity} $k$ is a (positive) scaling factor between the magnitudes of the tangent vectors across the boundary. A special case is found for $k = 1$, where the formulation leads to $C^1$ continuity and the Hermite polynomial. This parameter allows for flexibility in the mesh refinement.

\subsubsection{\texorpdfstring{Imposing $G^1$ on the field}{Imposing G¹ on the field}}

Now, we focus on how the geometrical tools derived so far can be used to interpolate a field. To better understand which information is needed on each node, it is better to shift the perspective from the element to a single vertex.

Following the derivation in \eqref{eq:derivates_in_00}, we know that the values and derivative in one of the corners only depends on the three closest control points. Given this, we define a local $3\times3$ grid of control points, denoted $P_{i,j}$ where $(i,j) \in \{-1, 0, 1\}^2$, centered on the shared vertex $P_{0,0}$. This 9-point grid is a subset of the control points from the four adjoining 16-point patches. For example, for one patch, $P_{0,0}$ corresponds to its corner control point, $P_{1,0}$ to its first tangential control point, and $P_{1,1}$ to its first twist control point.

The isoparametric property means that each control point is a vector $P_{i,j} = (R_{i,j}, Z_{i,j}, w_{i,j})$, where $(R, Z)$ are the spatial coordinates and $w$ is the field. We assume the mesh geometry $(R_{i,j}, Z_{i,j})$ is given. The unknowns are the nine scalar values $w_{i,j}$, which constitute the initial DOFs at the node $P_{0,0}$.

With this vertex-centered notation, the $G^1$ continuity condition \eqref{eq:g1_continuity} can be expressed more generally for the two principal directions s and t:
\begin{align}
P_{1,j} - P_{0,j} &= k_s (P_{0,j} - P_{-1,j}) \quad \forall j \in {-1, 0, 1} \\
P_{i,1} - P_{i,0} &= k_t (P_{i,0} - P_{i,-1}) \quad \forall i \in {-1, 0, 1}
\end{align}
where $k_s$ and $k_t$ are scalar constants determined by the mesh geometry.

Since our geometry is fixed, the equations can be used directly on $w_{i,j}$ to derive the necessary DoFs. Consider the equations for the $s$-direction:

\begin{align}
w_{1,j} - w_{0,j} &= k_s (w_{0,j} - w_{-1,j}) \quad \forall j \in {-1, 0, 1}
\label{eq:dof_on_s}
\end{align}

As \eqref{eq:dof_on_s} shows, the DoFs are not independent; they must lie on a line in the field space. 

We can now define the directed distances along the control polygon legs, from the central node as:
\begin{align}
d_{u,j} &= \text{sign}(j) \cdot ||M_{j,0} - M_{0,0}|| \quad \text{for } j = \pm 1 \\
d_{v,i} &= \text{sign}(i) \cdot ||M_{0,i} - M_{0,0}|| \quad \text{for } i = \pm 1
\end{align}
where $M_{i,j}$ are the projection on $(O,x,y)$:
\begin{equation}
M_{i,j} = (R_{i,j}, Z_{i,j})
\end{equation}
The values $d_{u,i}$ and $d_{v,j}$ are scalar properties derived from the mesh geometry.

The first two DoFs, after the field value $w_{0,0}$ itself, are $a_{0,0}$ and $b_{0,0}$, defined as:
\begin{align}
a_{0,0} \equiv \frac{w_{1,0} - w_{0,0}}{d_{u,1}} = \frac{w_{-1,0} - w_{0,0}}{d_{u,-1}} \\
b_{0,0} \equiv \frac{w_{0,1} - w_{0,0}}{d_{v,1}} = \frac{w_{0,-1} - w_{0,0}}{d_{v,-1}}.
\label{eq:ab00_definition}
\end{align}

Imposing the equivalence on the DoFs guarantees the required smoothness of the solution.

This handles the derivatives in $s$ and $t$, reducing the four DoFs related to the direct neighbors of the vertex to only two, but leaves us with the open issue of handling the mixed derivatives at the twist/corner points. 

For each of the four corners is possible to define:
\begin{equation}
    {\mathbf{m}}_{i,j} \equiv P_{i,j} + P_{0,0} - P_{i,0} - P_{0,j}.
\end{equation}

This serves the purpose of approximating a discrete mixed derivative $\frac{\partial^2 P}{\partial s \partial t}$; geometrically, this follows the parallelogram rule.

It is possible to prove that:
\begin{equation}
    \mathbf{m}_{i,j} = \frac{d_{u,i}}{d_{u,k}} {\mathbf{m}}_{k,j}, \qquad {\mathbf{m}}_{k,j} = \frac{d_{v,j}}{d_{v,l}} {\mathbf{m}}_{k,l}, \quad \forall i, j, k, l = \pm 1,
\end{equation}

Yielding the existence at the corner of a vector independent from the local twist points:
\begin{equation}
    \frac{1}{d_{u,i}d_{v,j}}\mathbf{m}_{i,j} = \frac{1}{d_{u,k}d_{v,l}}\mathbf{m}_{k,l} = \vec{\Gamma}, \quad \forall i, j, k, l = \pm 1.
\end{equation}

Therefore, calling $\gamma_{0,0}$ the third coordinate of $\vec{\Gamma}$, we obtain:
\begin{equation}
    \psi_{i,j} + \psi_{0,0} - \psi_{i,0} - \psi_{0,j} = \gamma_{0,0}d_{u,i}d_{v,j} \quad \forall i, j = \pm 1.
\end{equation}

Consequently, only four degrees of freedom ($\psi_{0,0}, \alpha_{0,0}, \beta_{0,0}, \gamma_{0,0}$), instead of nine, are necessary for each fluid quantity to determine all the control points associated with vertex $P_{0,0}$:
\begin{equation}
    \begin{cases}
        \psi_{i,0} = \psi_{0,0} + \alpha_{0,0}d_{u,i} \\
        \psi_{0,j} = \psi_{0,0} + \beta_{0,0}d_{v,j} \\
        \psi_{i,j} = \psi_{i,0} + \psi_{0,j} - \psi_{0,0} + \gamma_{0,0}d_{u,i}d_{v,j}
    \end{cases}
    \quad \forall i, j = \pm 1.
\end{equation}

\subsubsection{Field Interpolation}

A basis of normalized vectors is defined at a vertex \(P_{0,0}\). This basis consists of three vectors, \(\mathbf{U}_{\!0,0}\), \(\mathbf{V}_{\!0,0}\), and \(\mathbf{W}_{\!0,0}\):
\begin{equation}
\mathbf{U}_{\!0,0} = \begin{pmatrix} u_{0,0,x} \\ u_{0,0,y} \\ \alpha_{0,0} \end{pmatrix}, \quad
\mathbf{V}_{\!0,0} = \begin{pmatrix} v_{0,0,x} \\ v_{0,0,y} \\ \beta_{0,0} \end{pmatrix}, \quad
\mathbf{W}_{\!0,0} = \begin{pmatrix} w_{0,0,x} \\ w_{0,0,y} \\ \gamma_{0,0} \end{pmatrix}
\end{equation}

where the spatial subvectors \(\tilde{u}_{0,0}\), \(\tilde{v}_{0,0}\), and \(\tilde{w}_{0,0}\) (from Fig. 3.1(b)) are given by:
\begin{equation}
\begin{cases}
\tilde{u}_{0,0} = \frac{M_{1,0} - M_{0,0}}{d_{u,1}} \\
\tilde{v}_{0,0} = \frac{M_{0,1} - M_{0,0}}{d_{v,1}} \\
\tilde{w}_{0,0} = \frac{M_{1,1} + M_{0,0} - M_{0,1} - M_{1,0}}{d_{u,1}d_{v,1}}
\end{cases}
\end{equation}

The basis is written at each vertex, the Bézier parameterization of \(\mathbf{P} = (x, y, \psi)\) on element \(E_K\) can be rewritten as:
\begin{equation}
    \mathbf{P}(s,t) = \sum_{k=1}^{4} \tilde{\mathbf{P}}_{i_k}(s,t) = \sum_{i=1}^{4} 
    \begin{pmatrix} 
        \tilde{x}_{i_k}(s,t) \\ 
        \tilde{y}_{i_k}(s,t) \\ 
        \tilde{\psi}_{i_k}(s,t) 
    \end{pmatrix},
\end{equation}
where index \(i\) denotes vertex \(i\). The four components \(\tilde{\mathbf{P}}_i\) are given by:
\begin{equation}
    \left\{
    \begin{aligned}
        \tilde{\mathbf{P}}_{i_1}(s,t) &= (1-s)^2(1-t)^2[(1+2s)(1+2t)\mathbf{P}_{i_1} + 3s(1+2t)d_{u,i_1}\mathbf{U}_{i_1} \\
        &\qquad + 3t(1+2s)d_{v,i_1}\mathbf{V}_{i_1} + 9st d_{u,i_1}d_{v,i_1}\mathbf{W}_{i_1}], \\
        \\
        \tilde{\mathbf{P}}_{i_2}(s,t) &= s^2(1-t)^2[(3-2s)(1+2t)\mathbf{P}_{i_2} + 3(1-s)(1+2t)d_{u,i_2}\mathbf{U}_{i_2} \\
        &\qquad + 3t(3-2s)d_{v,i_2}\mathbf{V}_{i_2} + 9(1-s)t d_{u,i_2}d_{v,i_2}\mathbf{W}_{i_2}], \\
        \\
        \tilde{\mathbf{P}}_{i_3}(s,t) &= s^2t^2[(3-2s)(3-2t)\mathbf{P}_{i_3} + 3(1-s)(3-2t)d_{u,i_3}\mathbf{U}_{i_3} \\
        &\qquad + 3(3-2s)(1-t)d_{v,i_3}\mathbf{V}_{i_3} + 9(1-s)(1-t)d_{u,i_3}d_{v,i_3}\mathbf{W}_{i_3}], \\
        \\
        \tilde{\mathbf{P}}_{i_4}(s,t) &= (1-s)^2t^2[(1+2s)(3-2t)\mathbf{P}_{i_4} + 3s(3-2t)d_{u,i_4}\mathbf{U}_{i_4} \\
        &\qquad + 3(1+2s)(1-t)d_{v,i_4}\mathbf{V}_{i_4} + 9s(1-t)d_{u,i_4}d_{v,i_4}\mathbf{W}_{i_4}],
    \end{aligned}
    \right.
\end{equation}
Each of the (\(\tilde{\mathbf{P}}_i\)) can be set into the general form:
\begin{equation}
    \tilde{\mathbf{P}}_i(s,t) = b_{i,1|K}(s,t)\mathbf{P}_i + b_{i,2|K}(s,t)\mathbf{U}_i + b_{i,3|K}(s,t)\mathbf{V}_i + b_{i,4|K}(s,t)\mathbf{W}_i,
\end{equation}

$b_{i,j}^{(K)}(s,t)$ are the Bézier basis functions.

Elements with a higher order of continuity are also available in the JOREK code, although not considered in this work. A more detailed discussion on geometrical continuity of Bézier patches can be found in \cite{GN_Bezier}.

\subsubsection{Bounding Box of a Bicubic Bézier Element}
\label{sec:AABB_bezier}

As we will see in section \ref{sec:rtree}, for some algorithms it is important to know the bounding box of each element. The bounding box of an element is defined as the smallest, axis-aligned rectangle that contains all the points belonging to an element. 

Let's start by considering the physical coordinates $(R, Z)$, at a specific toroidal angle $\phi$ they are a function of the local coordinates $(s, t) \in [0, 1]^2$:
\begin{equation}
\mathbf{x}(s,t) = \begin{pmatrix} R(s,t) \\ Z(s,t) \end{pmatrix} = \sum_{i=0}^{3} \sum_{j=0}^{3} \mathbf{P}_{i,j}(\phi) \, B^{3}_{i}(s) B^{3}_{j}(t)
\end{equation}

Finding the extreme points of $R(s,t)$ and $Z(s,t)$ would require solving $\nabla R(s,t) = 0$ and $\nabla Z(s,t) = 0$ over the domain $(s,t) \in [0, 1]^2$, an expensive problem to solve for each element. A robust and efficient approximation consists of finding the extrema on the boundaries of the element. This assumption holds for the meshes taken into consideration, and is enforced at the start of the simulation by assuring that the Jacobian determinant does not change sign inside the element.

The search for extrema is restricted to the four edges of the element: $s=0, s=1, t=0,$ and $t=1$. Consider the coordinate $R$ along the edge $t=0$, for $s \in [0, 1]$. The function $R(s, t=0)$ simplifies to a 1D cubic Bézier curve:
\begin{equation}
R(s, t=0) = \sum_{i=0}^{3} R_{i,0}(\phi) \, B^{3}_{i}(s)
\end{equation}
This is a cubic polynomial in $s$. To find its local extrema for $s \in (0, 1)$, we must find the roots of its derivative with respect to $s$:
\begin{equation}
\frac{d}{ds} R(s, t=0) = \frac{d}{ds} \left( \sum_{i=0}^{3} R_{i,0}(\phi) \, B^{3}_{i}(s) \right) = 0
\end{equation}
Since the Bernstein polynomials $B_i^3(s)$ are cubic in $s$, their derivative is quadratic. This results in a simple quadratic equation of the form $as^2 + bs + c = 0$, which can be efficiently solved for each boundary, for each dimension. 

By taking the minimum and maximum along each boundary, we get the corners of the bounding box for each element.

\subsubsection{Toroidal Fourier expansion}

To account for the 3D nature of the problem, we need to discretize along the toroidal direction, especially in non-axisymmetric devices such as stellarators. Every physical or geometric quantity can be expressed in a Fourier series of the toroidal angle $\Phi$. Since everything we are interested in is real-valued, we use a basis consisting of real trigonometric functions. Any quantity $X$ is expressed as:
\begin{equation}
X(s, t, \Phi) = \sum_{l=1}^{n_{\text{tor}}} X_l(s,t) \, Z_l(\Phi)
\end{equation}
where $X_l$ are the real-valued Fourier coefficients (which depend on the poloidal coordinates), and the basis functions $Z_l(\Phi)$ are defined in \cite{rac_gpu} as:
\begin{equation}
Z_l(\Phi) =
\begin{cases}
    1 & \text{if } l=1 \quad (\text{for the } n=0 \text{ mode}) \\
    \cos\left(n_{\text{period}} \frac{l}{2} \Phi\right) & \text{if } l > 1 \text{ is even} \\
    \sin\left(n_{\text{period}} \frac{l-1}{2} \Phi\right) & \text{if } l > 1 \text{ is odd}
\end{cases}
\end{equation}
Here, $n_{\text{period}}$ denotes the field-periodicity of the configuration, allowing for representations of geometries that complete their pattern multiple times in one full $2\pi$ turn.

\subsubsection{The Finite Element Method}

Now, we have all the tools necessary to build the numerical framework to solve the governing equations that describe the plasma.

To accomplish this task, we use the Finite Element Method \cite{FEM}. By subdividing the domain into smaller subdomains, looking for a suitable coefficient, we will be able to numerically approximate the function.

In our case, the poloidal domain $\mathcal{D}$ is partitioned into a mesh of non-overlapping Bézier patch finite elements $\mathcal{D}_K$, such that:

\begin{equation}
\mathcal{D} = \bigcup_K \mathcal{D}_K
\end{equation}

For each element $K$, we associate a set of nodes where the degrees of freedom are defined. In addition to the vertices, each element carries four additional control points that influence the shape of the basis functions within the patch. The local basis functions are $b_{i,j}^{(K)}(s,t)$, where $(s,t)$ are local coordinates within the element $\mathcal{D}_K$, and the indices $i,j$ represent the nodes and control points associated with it.

The solution within each element is expressed as a linear combination of these local basis functions:

\begin{equation}
\psi^{(K)}(s, t) = \sum_{i,j} \psi_{i,j}^{(K)} \, b_{i,j}^{(K)}(s, t)
\end{equation}

As previously noted, the solution needs to be three-dimensional, so we take that into account by introducing the Fourier expansion along the toroidal angle $\Phi$:

\begin{equation}
\psi^{(K)}(s, t, \Phi) = \sum_{l=1}^{n_{\text{tor}}} \sum_{i,j} \psi_{i,j,l}^{(K)} \, b_{i,j}^{(K)}(s,t) \, Z_l(\Phi)
\label{field_expansion}
\end{equation}

Where:
\begin{itemize}
    \item $Z_l(\Phi)$ are the toroidal basis functions
    \item $\psi_{i,j,l}^{(K)}$ are the local degrees of freedom associated with the $l$-th toroidal mode and the $(i,j)$-th local basis function within element $K$,
    \item $b_{i,j}^{(K)}(s,t)$ are the Bézier basis functions defined locally over element $\mathcal{D}_K$.
\end{itemize}

\subsection{Governing equations and Numerical Approximations for the kinetics}

In this section, we are going to provide a general overview of the kinetics equations and the numerical approximations considered, as well as their stability. For a detailed description, see \cite{Boris_Mover}.

\subsubsection{Newton-Lorentz equation}

We study the motion of charged particles in a given electromagnetic field, according to Newton's equations:

\begin{equation}
\frac{d\mathbf{x}}{dt} = \mathbf{v},
\label{eq:position}
\end{equation}

\begin{equation}
m\frac{d\mathbf{v}}{dt} = q[\mathbf{E}(\mathbf{x}) + \mathbf{v} \times \mathbf{B}(\mathbf{x})],
\label{eq:velocity}
\end{equation}

Where $\mathbf{x}$ ($\mathbf{v}$) is the position (velocity) of the particle with mass $m$ and charge $q$, while $\mathbf{E}$ ($\mathbf{B}$) is the given electric (magnetic) field.

The electromagnetic field is always considered given, as it is calculated by solving the MHD equations before starting the kinetic solver, and is considered a static field during the particle pushing.

\subsubsection{The Boris Pusher}

For the case under study, we chose to employ an explicit time stepping scheme called the Boris pusher, an integration scheme with second order accuracy and comparably light computational requirements.

This algorithm is considered a leap-frog type of algorithm, since it employs staggered updates of position and velocity in time. In particular,
velocities are updated at time $t = t^{n+1/2} = (n + \frac{1}{2})\Delta t$ and position are updated like $t = t^n = n\Delta t$.

The following discussion stems from \cite{Boris_Mover}, which presents a detailed analysis of the numerical properties of such algorithms, out of scope for the discussion carried out in this work.

The algorithm separates the two different equations, the electric and the magnetic parts of the Lorentz force, by performing the following three steps:

A half-step update (time step $\Delta t/2$) with the contribution of only the electric field:
\begin{equation}
\mathbf{v}^- = \mathbf{v}^{n-1/2} + \frac{q}{m}\mathbf{E}(\mathbf{x}^n)\frac{\Delta t}{2}. 
\label{eq:half_step_electric1}
\end{equation}

Perform a full step (time step $\Delta t$) following the magnetic field:
\begin{equation}
\mathbf{v}' = \mathbf{v}^- + f^{n,\Delta t}\mathbf{v}^- \times \mathbf{B}(\mathbf{x}^n),
\label{eq:full_step_magnetic1}
\end{equation}

\begin{equation}
\mathbf{v}^+ = \mathbf{v}^- + \frac{2f^{n,\Delta t}}{1 + (f^{n,\Delta t})^2\mathbf{B}(\mathbf{x}^n)^2}\mathbf{v}' \times \mathbf{B}(\mathbf{x}^n),
\label{eq:full_step_magnetic2}
\end{equation}

with
\begin{equation}
f^{n,\Delta t} = \frac{\tan(\frac{q}{m}\frac{\Delta t}{2}|\mathbf{B}(\mathbf{x}^n)|)}{|\mathbf{B}(\mathbf{x}^n)|}.
\label{eq:f_parameter}
\end{equation}

Perform another half-step update (time step $\Delta t/2$) with acceleration only due to the electric field:
\begin{equation}
\mathbf{v}^{n+1/2} = \mathbf{v}^+ + \frac{q}{m}\mathbf{E}(\mathbf{x}^n)\frac{\Delta t}{2}.
\label{eq:half_step_electric2}
\end{equation}

Once the velocity update is completed, the position can be updated  by
\begin{equation}
\mathbf{x}^{n+1} = \mathbf{x}^n + \mathbf{v}^{n+1/2}\Delta t.
\label{eq:position_update}
\end{equation}

This algorithm is second-order accurate and time reversible \cite{Boris_Mover}. Of course, as for any leap-frog algorithm, the accuracy also depends on the choice of the initial step that needs to be provided to relate the (known) particle position and velocity at time $t = 0$ to the velocity $\mathbf{v}^{-1/2}$ (or $\mathbf{v}^{1/2}$) that the algorithm needs to get started.

\subsubsection{Projecting particles on the Finite Element grid}

To couple the kinetic particles with the fluid part, it is necessary to translate the discrete information from the particles into a continuous field compatible with our Finite Element formulation of the MHD equations. We employ a projection using the previously established methodologies. The goal is to find a continuous field, $p$, that is equivalent in a weighted-average sense to the source data from the particles, $X$.

This equivalence is formally expressed in the weak form, which states that the two quantities are identical when integrated against any valid test function $v$:
\begin{equation}
    \int_V p v \, dV = \int_V X v \, dV
\end{equation}

Here, $p$ is the projected field to be solved for (e.g., mass density, parallel velocity), and $X$ represents the corresponding moment of the particle distribution function, 1 for number density, $L_z$ for radiative energy loss, $m$ for mass density, $v_\parallel$ for parallel velocity, $E$ for energy et cetera \cite{DanThesis}.

Usually, in these particle methods, we introduce a shape function $S(x)$ describing the shape of the particle in space. Various shape functions can be chosen and consequently give rise to different types of smoothing effects on the particle; by including these shape functions into the formulation, we get:

\begin{equation}
    \int_V p v \, dV = \int_V \left( \sum_{i=1}^{N} S(\mathbf{x} - \mathbf{x}_i)w_i  X  \right) v \, dV
    \label{weak_form}
\end{equation}
The term $w_i$ indicates the weight we give to each particle in the simulation. This is done to create macroparticles, whose evolution is tied together but which carry different amounts of physical properties.

\paragraph{Matrix construction}\

Let's now look at the single element to construct the matrix representing the linear system of equations that needs to be solved for the projections. Within each element $\mathcal{D}_K$, the field $p$, following the formulation introduced in \eqref{field_expansion}, is approximated as:
\begin{equation}
p^{(K)}(s, t, \Phi) = \sum_{l=1}^{n_{\text{tor}}} \sum_{i,j} p_{i,j,l}^{(K)} \, b_{i,j}^{(K)}(s,t) \, Z_l(\Phi)
\end{equation}

Similarly, the test function $v$ is expressed as:
\begin{equation}
v^{(K)}(s, t, \Phi) = \sum_{m=1}^{n_{\text{tor}}} \sum_{k,\ell} v_{k,\ell,m}^{(K)} \, b_{k,\ell}^{(K)}(s,t) \, Z_m(\Phi)
\end{equation}

The mapping from local coordinates $(s, t, \Phi)$ to physical cylindrical coordinates $(R, Z, \phi)$ is given by:
\begin{equation}
(s, t, \Phi) \mapsto \mathbf{x}(s, t, \Phi) = \left( R(s, t, \Phi), \; Z(s, t, \Phi), \; \phi=\Phi \right)
\end{equation}

The Jacobian of the 2D poloidal mapping $(s,t) \mapsto (R,Z)$ is:
\begin{equation}
J_{\text{pol}}^{(K)}(s, t, \Phi) = 
\det \begin{bmatrix}
\frac{\partial R}{\partial s} & \frac{\partial R}{\partial t} \\
\frac{\partial Z}{\partial s} & \frac{\partial Z}{\partial t}
\end{bmatrix}
\label{eq:jacobian}
\end{equation}

Note that $J_{\text{pol}}^{(K)}$ depends in general on all three coordinates $s$, $t$, and $\Phi$ due to the dependence of $R$ and $Z$. The physical volume element $dV = R \, dR \, dZ \, d\phi$ transforms to the local coordinates as:
\begin{equation}
dV = R(s,t,\Phi) \, J_{\text{pol}}^{(K)}(s,t,\Phi) \, ds \, dt \, d\Phi =: \mathcal{J}^{(K)}(s,t,\Phi) \, ds \, dt \, d\Phi
\end{equation}
where $\mathcal{J}^{(K)} = R \cdot J_{\text{pol}}^{(K)}$ is the determinant of the full 3D coordinate transformation Jacobian.

The element-wise contribution to the left-hand side of equation \eqref{weak_form} is:
\begin{equation}
\int_{\mathcal{D}_K} \int_0^{2\pi} p^{(K)}(s, t, \Phi) \, v^{(K)}(s, t, \Phi) \, \mathcal{J}^{(K)}(s, t, \Phi) \, d\Phi \, ds \, dt
\end{equation}

Substituting the basis function expansions for $p^{(K)}$ and $v^{(K)}$ gives the element mass matrix contribution. The equation for an arbitrary test function $v$ is satisfied if it holds for each basis function. This leads to a linear system where the element mass matrix $M^{(K)}$ couples the degrees of freedom $p_{i,j,l}^{(K)}$. A single entry of this matrix, which couples the poloidal basis functions indexed by $(i,j)$ and $(k,\ell)$ and the toroidal modes $l$ and $m$, is given by:
\begin{equation}
M_{ij,k\ell,lm}^{(K)} = \int_0^{2\pi} \left[ \int_{\mathcal{D}_K} b_{i,j}^{(K)}(s,t) \, b_{k,\ell}^{(K)}(s,t) \, \mathcal{J}^{(K)}(s, t, \Phi) \, ds \, dt \right] Z_l(\Phi) \, Z_m(\Phi) \, d\Phi
\label{mat_contribution}
\end{equation}

To evaluate this integral numerically, we can conceptually separate the integration over the poloidal plane from the integration over the toroidal angle. Let's define an intermediate function, $F_{ij,k\ell}(\Phi)$, which represents the result of the poloidal integral for a given toroidal angle $\Phi$:
\begin{equation}
F_{ij,k\ell}(\Phi) := \int_{\mathcal{D}_K} b_{i,j}^{(K)}(s,t) \, b_{k,\ell}^{(K)}(s,t) \, \mathcal{J}^{(K)}(s, t, \Phi) \, ds \, dt
\end{equation}
The integral over the 2D reference element $\mathcal{D}_K$ is approximated using a fourth-order Gaussian quadrature rule. This is constructed as the product of two 1D rules, leading to a double summation over the quadrature points $(s_{ms}, t_{mt})$ with their corresponding 1D weights $\omega_{ms}$ and $\omega_{mt}$:
\begin{equation}
F_{ij,k\ell}(\Phi) \approx \sum_{ms=1}^{n_{\text{gp}}} \sum_{mt=1}^{n_{\text{gp}}} \omega_{ms} \omega_{mt} \, b_{i,j}^{(K)}(s_{ms}, t_{mt}) \, b_{k,\ell}^{(K)}(s_{ms}, t_{mt}) \, \mathcal{J}^{(K)}(s_{ms}, t_{mt}, \Phi)
\end{equation}

With this intermediate function, the expression for the mass matrix entry simplifies to a single integral over the toroidal angle:
\begin{equation}
M_{ij,k\ell,lm}^{(K)} = \int_0^{2\pi} F_{ij,k\ell}(\Phi) \, Z_l(\Phi) \, Z_m(\Phi) \, d\Phi
\label{eq:coupled_harmonics}
\end{equation}
Since the integral in \eqref{eq:coupled_harmonics} is integrating a periodic function, it can be evaluated using a periodic Gaussian quadrature, which is exact with $N$ points for trigonometric polynomials of degree $2N-2$ \cite{period_gauss_quadrature}. We therefore sample the function in $n_{plane}$ equidistant points along the toroidal direction.
\begin{equation}
M_{ij,k\ell,lm}^{(K)} = \frac{2\pi}{n_{plane}}\sum_{p=1}^{n_{\text{plane}}} \, F_{ij,k\ell}(\Phi_p) \, Z_l(\Phi_p) \, Z_m(\Phi_p)
\label{eq:coupled_matrix}
\end{equation}

\paragraph{Uncoupled Harmonics in an Axisymmetric Geometry}\

The coupling between different toroidal harmonics on the left-hand side comes directly from the geometry of the mesh, see eq. \eqref{mat_contribution}.
The term responsible for coupling the toroidal modes $l$ and $m$ is the full 3D Jacobian determinant, $\mathcal{J}^{(K)}(s, t, \Phi) = R(s,t,\Phi) \, J_{\text{pol}}^{(K)}(s,t,\Phi)$. If this Jacobian is a function of the toroidal angle $\Phi$, the integral over $\Phi$ will generally be different from zero for $l \neq m$.

An important simplification happens if the mesh grid is axisymmetric. In an axisymmetric grid, the poloidal section is the same, independently of toroidal angle $\Phi$. This means the mapping from the local coordinates $(s,t)$ to the physical poloidal coordinates $(R,Z)$ is independent of $\Phi$:
\begin{equation}
(s, t) \mapsto \mathbf{x}_{\text{pol}}(s, t) = \left( R(s, t), \; Z(s, t) \right) \quad (\text{independent of } \Phi)
\end{equation}
Therefore, the derivatives are also independent of $\Phi$. This directly changes the Jacobian in the poloidal section by providing independence from the toroidal direction.

With this simplification, the mass matrix integral can be rearranged, as the term in the square brackets is no longer a function of $\Phi$:
\begin{equation}
M_{ij,k\ell,lm}^{(K)} = \left[ \int_{\mathcal{D}_K} b_{i,j}^{(K)}(s,t) \, b_{k,\ell}^{(K)}(s,t) \, \mathcal{J}^{(K)}(s, t) \, ds \, dt \right] \left[ \int_0^{2\pi} Z_l(\Phi) \, Z_m(\Phi) \, d\Phi \right]
\end{equation}
The basis set $\{Z_l(\Phi)\}$ is constructed from sines, cosines, and a constant term, which are orthogonal over a period. This means the integral is zero unless the two basis functions are the same:
\begin{equation}
\int_0^{2\pi} Z_l(\Phi) \, Z_m(\Phi) \, d\Phi = C_l \, \delta_{lm}
\end{equation}

where $C_l$ is the normalization constant given by
\begin{equation}
C_l =
\begin{cases}
    \displaystyle 2\pi & \text{if } l=1 \\
    \displaystyle \pi & \text{if } l > 1
\end{cases}
\end{equation}

and $\delta_{lm}$ is the Kronecker delta.

The appearance of the delta means that the matrix entry $M_{ij,k\ell,lm}^{(K)}$ is different from zero only if the toroidal mode indices are identical ($l=m$). Therefore, for an axisymmetric grid like the already existing projections for tokamak cases in the code, the global mass matrix is perfectly block-diagonal when rearranged to cluster the harmonics contribution together. 

Each harmonic is decoupled, allowing the full linear system to be solved as a set of smaller, independent systems. A similar scheme is employed in the iterative solver of the fluid as an effective and efficient preconditioner for the fluid matrix \cite{UncoupledPreconditioner, ASTER_Iterative}.

This is not the case for stellarator scenarios considered in this work.

\paragraph{Fast Fourier Transform for the evaluation of the toroidal integral}\

\label{sec:fft_math}

For stellarator simulations, a high number of harmonics is usually needed to accurately represent solutions on the plasma, given the intrinsic coupling of harmonics due to the non-axisymmetric geometry; the matrix assembly becomes a bottleneck for simulations. However, there is a computationally more efficient algorithm, used in the fluid part, to evaluate the toroidal integral by using the Fast Fourier Transform \cite{CooleyTukey}.

Let's define the integer wavenumber $\nu$:
\begin{equation}
  \nu(l)\;=\;\bigl\lfloor l/2\bigr\rfloor
  \quad\Longrightarrow\quad
  \nu(1)=0,\,
  \nu(2)=\nu(3)=1,\,
  \nu(4)=\nu(5)=2,\dots
  \label{eq:k_integer_bis}
\end{equation}

Such that the couple of $sin$ and $cos$ with the same periodicity can be addressed by the same wavenumber.

With these identities, every real basis function can be rewritten as
\begin{equation}
  Z_{2q}(\Phi)=\cos(q\Phi)
               =\tfrac12\bigl(
                   e^{\mathrm i q\Phi}+e^{-\mathrm i q\Phi}\bigr),
  \qquad
  Z_{2q+1}(\Phi)=\sin(q\Phi)
               =\tfrac1{2\mathrm i}\bigl(
                   e^{\mathrm i q\Phi}-e^{-\mathrm i q\Phi}\bigr).
\end{equation}

Therefore, any product $Z_l\,Z_m$ can be written as a linear combination of two complex exponentials:

\begin{equation}
  Z_l(\Phi)\,Z_m(\Phi)
  \;=\;
  \tfrac12\Bigl(
      e^{\mathrm i[\nu(l)-\nu(m)]\Phi}
     +\sigma_{lm}\,
       e^{\mathrm i[\nu(l)+\nu(m)]\Phi}
     \Bigr)
  \label{eq:prod_identity_bis}
\end{equation}

where the coefficient \(\sigma_{lm}\) depends only on the parity of the two indices:
\begin{equation}
  \sigma_{lm}
  \;=\;
  \begin{cases}
      +1 & \text{$l$ and $m$ \emph{have the same parity}} 
               \;\;\bigl[\cos\!\times\!\cos \;\text{or}\; \sin\!\times\!\sin\bigr],\\[6pt]
      -1 & \text{$l$ and $m$ \emph{have opposite parity}}
               \;\;\;\bigl[\cos\!\times\!\sin \;\text{or}\; \sin\!\times\!\cos\bigr].
  \end{cases}
  \label{eq:sigma_parity}
\end{equation}

The two exponential factors carry wavenumbers
\begin{equation}
  r_1 := \nu(l)-\nu(m), \qquad
  r_2 := \nu(l)+\nu(m),
\end{equation}
both of which obey $|r_{1,2}|<n_{\mathrm{plane}}/2$ in practice, because we strictly choose
$n_{\mathrm{plane}}\ge 2\,n_{\mathrm{tor}}$
in the simulations.

Inserting \eqref{eq:prod_identity_bis} into \eqref{eq:coupled_matrix} and distributing the summation gives:

\begin{equation}
\label{eq:M_midstep}
\begin{split}
  M_{ij,k\ell,lm}^{(K)}
  &= \frac{\pi}{n_{\mathrm{plane}}}
    \sum_{p=1}^{n_{\text{plane}}} F_{ij,k\ell}(\Phi_p)
       \left[
         e^{(\mathrm i r_1\Phi_p)}
        + \sigma_{lm}
          e^{(\mathrm i r_2\Phi_p)}
       \right] \\
  &= \frac{\pi}{n_{\mathrm{plane}}} \left[
    \sum_{p=1}^{n_{\text{plane}}} F_{ij,k\ell}(\Phi_p)e^{(\mathrm i r_1\Phi_p)} + \sigma_{lm} \sum_{p=1}^{n_{\text{plane}}} F_{ij,k\ell}(\Phi_p)e^{(\mathrm i r_2\Phi_p)}\right]
\end{split}
\end{equation}

We can now define the complex discrete Fourier coefficients:

\begin{equation}
  \widehat{F}_{ij,k\ell}^r
  :=\
  \sum_{p=1}^{n_{\mathrm{plane}}}
        F_{ij,k\ell}(\Phi_p)\;
        e^{-\mathrm i\,r\,\Phi_p},
  \qquad
  r = 1,\dots ,n_{\mathrm{plane}} .
  \label{eq:DFT_F}
\end{equation}

and by substituting them into \eqref{eq:DFT_F} we obtain:

\begin{equation}
\label{eq:M_final}
M_{ij,k\ell,lm}^{(K)} = \frac{\pi}{n_{\mathrm{plane}}} \left[ \widehat{F}_{ij,k\ell}^{-r_1} + \sigma_{lm} \widehat{F}_{ij,k\ell}^{-r_2} \right]
\end{equation}

Since all the coefficients can be obtained from a single FFT computation, the asymptotic complexity is greatly reduced. 

When using the direct summation, for each degree of freedom on the poloidal section, we have to evaluate the function in $n_{plane}$ different points for each couple $l,m$ of toroidal harmonics: 
\begin{equation}
  \text{cost}_{\text{direct}} = n_{\mathrm{tor}}^2 \times O(n_{\mathrm{plane}}) = O(n_{\mathrm{tor}}^2 \, n_{\mathrm{plane}}).
\end{equation}

On the other hand, using the FFT acceleration, we compute all the coefficients in a single FFT pass of complexity $O(n_{plane}log(n_{plane}))$. Once all the coefficients are obtained, we construct the matrix by carefully adding the relevant coefficients for each couple following \eqref{eq:prod_identity_bis}, resulting in a final complexity of:
\begin{equation}
\text{cost}_{\text{FFT}} = O(n_{\mathrm{plane}} \log n_{\mathrm{plane}} + n_{\mathrm{tor}}^2).
\end{equation}  

In the direct method, the expensive $O(n_{\mathrm{plane}})$ summation is repeated $N_{\mathrm{tor}}^2$ times. In the FFT method, the expensive part, with a slightly higher cost of $O(n_{\mathrm{plane}} \log n_{\mathrm{plane}})$, is performed only once. The subsequent assembly is extremely fast.

\paragraph{Right hand side}\

We consider now the right-hand side of the weak formulation from \eqref{weak_form}:
\begin{equation}
\int_V \left( \sum_{i=1}^{N} S(\mathbf{x} - \mathbf{x}_i) \, w_i \, X(\mathbf{x}) \right) v(\mathbf{x}) \, dV
\end{equation}
The element-wise contribution is obtained by substituting the expansion for $v$ and the transformation for $dV$:
\begin{equation}
\sum_{m=1}^{n_{\text{tor}}} \sum_{k,\ell} v_{k,\ell,m}^{(K)} \int_{\mathcal{D}_K} \int_0^{2\pi} \left( \sum_{i=1}^{N} S(\mathbf{x}(s,t,\Phi) - \mathbf{x}_i) \, w_i \, X(\mathbf{x}(s,t,\Phi)) \right) \, b_{k,\ell}^{(K)}(s,t) \, Z_m(\Phi) \, \mathcal{J}^{(K)}(s,t,\Phi) \, d\Phi \, ds \, dt
\end{equation}

We define the right-hand-side vector coefficients $L_{k,\ell,m}^{(K)}$ from the integral term as:
\begin{equation}
\label{eq:rhs_integral_general}
L_{k,\ell,m}^{(K)} = \int_{\mathcal{D}_K} \int_0^{2\pi} \left( \sum_{i=1}^{N} S(\mathbf{x}(s,t,\Phi) - \mathbf{x}_i) \, w_i \, X(\mathbf{x}_i) \right) \, b_{k,\ell}^{(K)}(s,t) \, Z_m(\Phi) \, \mathcal{J}^{(K)}(s,t,\Phi) \, d\Phi \, ds \, dt
\end{equation}
where for brevity we use $X(\mathbf{x}_i)$ as the value of the moment for particle $i$.

The evaluation of this integral depends on the choice of the shape function $S$. The simplest and most computationally efficient choice is the Dirac delta function, $S(\mathbf{x} - \mathbf{x}_i) = \delta(\mathbf{x} - \mathbf{x}_i)$. Using this choice, the integral in equation \eqref{eq:rhs_integral_general} can be greatly simplified.

The Dirac delta function transforms between coordinate systems according to the rule:
\begin{equation}
\delta(\mathbf{x} - \mathbf{x}_i) = \frac{\delta(s - s_i)\delta(t - t_i)\delta(\Phi - \Phi_i)}{|\mathcal{J}^{(K)}(s_i, t_i, \Phi_i)|}
\end{equation}
where $(s_i, t_i, \Phi_i)$ are the local coordinates of the particle at position $\mathbf{x}_i$. This transformation is valid only for particles located within the element $K$; otherwise, the delta function is zero over the integration domain $\mathcal{D}_K \times [0, 2\pi)$.

Substituting this into the expression for $L_{k,\ell,m}^{(K)}$ and swapping the sum and integral, we get:
\begin{equation}
L_{k,\ell,m}^{(K)} = \sum_{i \in K} \int_{\mathcal{D}_K} \int_0^{2\pi} \frac{\delta(s-s_i)\delta(t-t_i)\delta(\Phi-\Phi_i)}{|\mathcal{J}^{(K)}(s_i,t_i,\Phi_i)|} w_i X(\mathbf{x}_i) \, b_{k,\ell}^{(K)}(s,t) \, Z_m(\Phi) \, \mathcal{J}^{(K)}(s,t,\Phi) \, d\Phi \, ds \, dt
\end{equation}
The summation is now restricted to particles $i \in K$ that reside within the element $K$. By the sifting property of the delta function, the integral collapses to an evaluation of the integrand at the particle's local coordinates $(s_i, t_i, \Phi_i)$:
\begin{equation}
L_{k,\ell,m}^{(K)} = \sum_{i \in K} w_i X(\mathbf{x}_i) \, b_{k,\ell}^{(K)}(s_i,t_i) \, Z_m(\Phi_i) \, \frac{\mathcal{J}^{(K)}(s_i,t_i,\Phi_i)}{|\mathcal{J}^{(K)}(s_i,t_i,\Phi_i)|}
\end{equation}
Assuming the element mapping is not inverted, the Jacobian determinant $\mathcal{J}^{(K)}$ is positive, and the fraction equals one. This reduces the computationally expensive triple integral to a simple sum over the particles within the element:
\begin{equation}
L_{k,\ell,m}^{(K)} = \sum_{i \in K} w_i \, X(\mathbf{x}_i) \, b_{k,\ell}^{(K)}(s_i,t_i) \, Z_m(\Phi_i)
\end{equation}
This formulation provides a significant computational advantage, as it avoids numerical quadrature and only requires finding which particles reside in which element and evaluating the basis functions at the particle locations.

So, the element-wise right-hand side of the weak form becomes:
\begin{equation}
\sum_{m=1}^{n_{\text{tor}}} \sum_{k,\ell} v_{k,\ell,m}^{(K)} \left( \sum_{i \in K} w_i \, X(\mathbf{x}_i) \, b_{k,\ell}^{(K)}(s_i,t_i) \, Z_m(\Phi_i) \right)
\end{equation}


\section{Algorithms and verification for PiC methods in the JOREK stellarator model} 

\label{Chapter3} 

The new particle-in-cell (PiC) capabilities within the JOREK stellarator model are implemented and verified in this section after the theoretical underpinnings were established in the previous section. Here, the development and verification of the algorithms needed for a fully functional hybrid simulation are the main priorities. In order to verify that the basic particle tracing operates as anticipated in both tokamak and stellarator configurations, we start by presenting initial tests. Expanding upon this, we describe the main contribution of this work: the application of a fully coupled harmonic projection scheme, which is necessary to accurately depict fields in stellarator geometries that are not axisymmetric. First, qualitative projections onto realistic stellarator meshes are used to illustrate the efficacy of this approach, which is then quantitatively validated through a series of convergence tests. Lastly, we implement and evaluate a spatial R-Tree data structure to tackle the problem of effectively locating particles within the 3D mesh.

\subsection{Preliminary tests}

\subsubsection{Particle Tracing}

In order to produce a working hybrid solver for the stellarator model, it is crucial to have consistent kinetic simulation results with the existing code base. For this purpose, three simulations have been set up and tested in which particles are pushed in static fields: a tokamak configuration in a stellarator model, a tokamak configuration in a tokamak model, and a stellarator configuration in a stellarator model. The first and the second should have an agreeing result, while the third should showcase the non-axisymmetric behavior intrinsic to the stellarator geometry.

The stellarator model used is the model 180, while the tokamak model used is 600

\begin{figure}[!htbp]
    \centering
    \begin{minipage}[b]{0.48\textwidth}
        \centering
        \includegraphics[width=\textwidth]{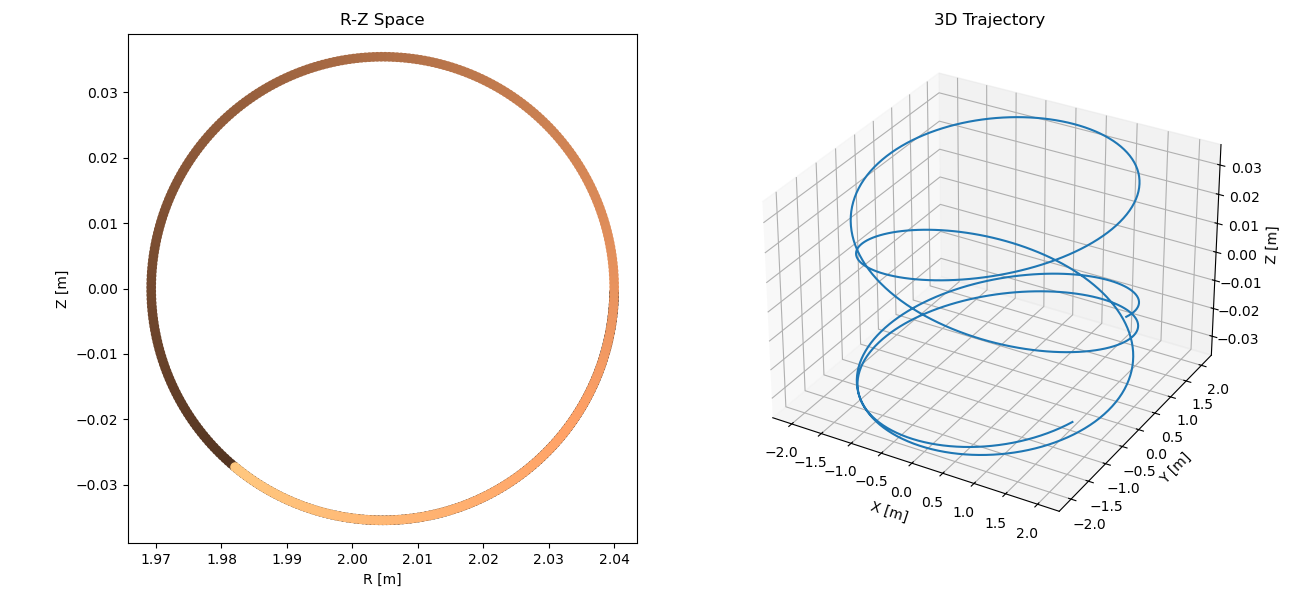}
        \label{fig:stell_tok_config}
    \end{minipage}
    \hfill
    \begin{minipage}[b]{0.48\textwidth}
        \centering
        \includegraphics[width=\textwidth]{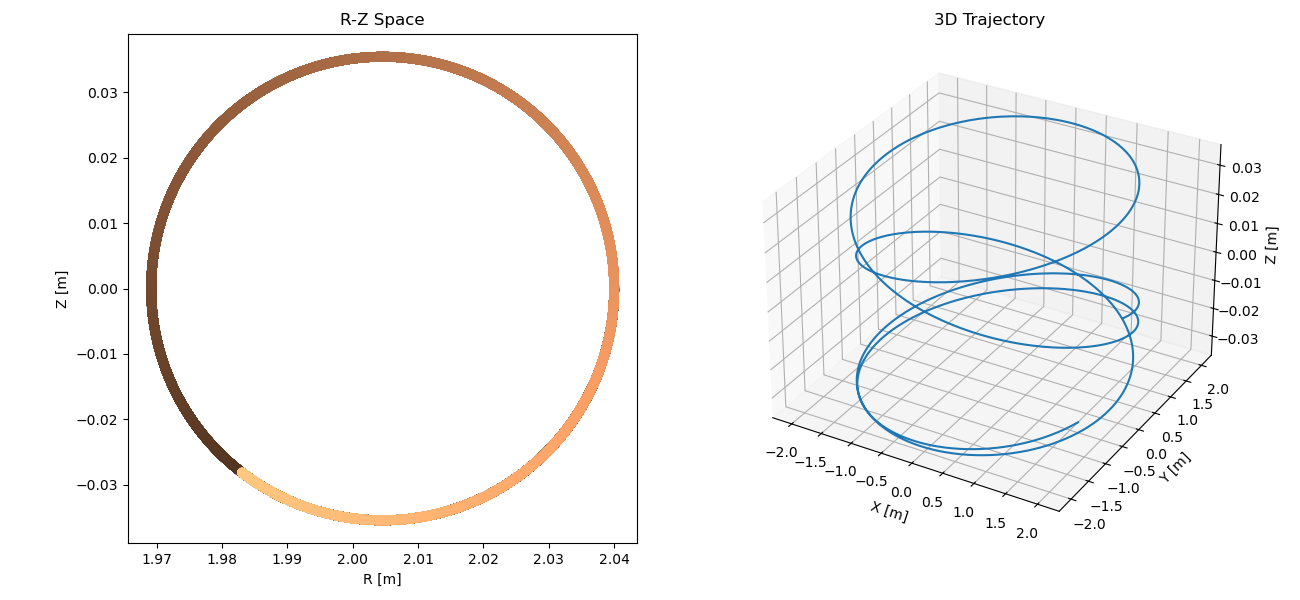}
        \label{fig:tok_tok_config}
    \end{minipage}
    \caption{Comparison of particle trajectories between stellarator and tokamak models in tokamak configuration.}
    \label{fig:trajectory_comparison}
\end{figure}

\begin{figure}[H]
    \centering
    \includegraphics[width=0.75\linewidth]{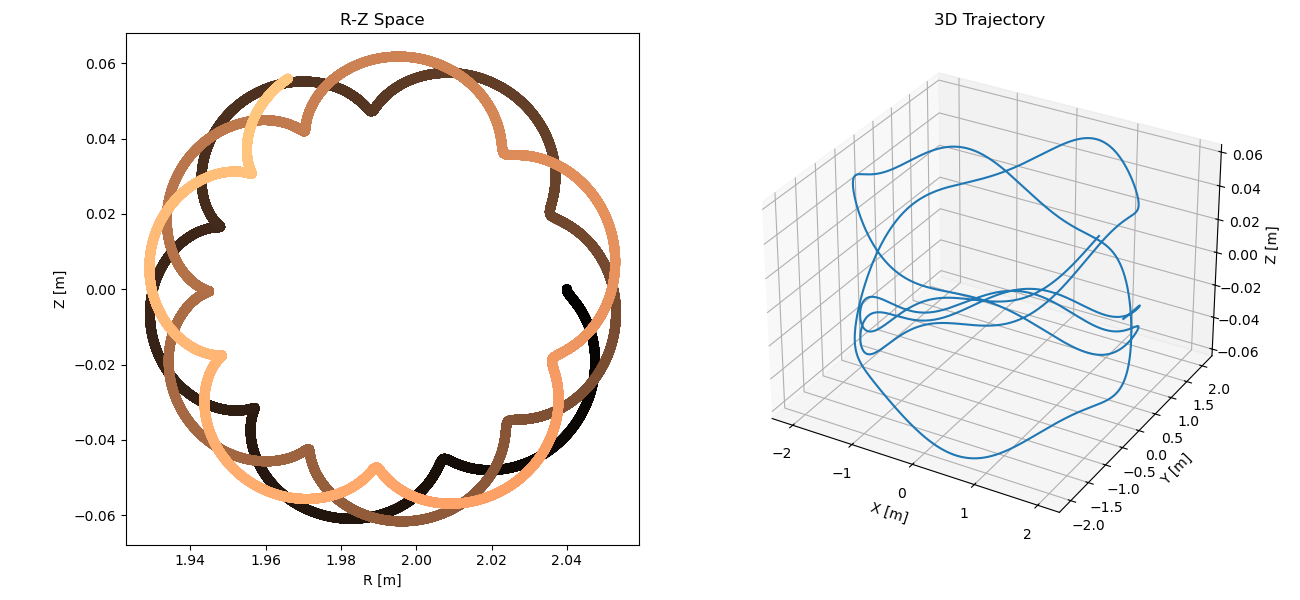}
    \caption{Particle Tracing in a stellarator case using the stellarator model}
    \label{fig:true_stel}
\end{figure}

As expected, the first two configurations yield the same result within numerical tolerance as seen in Fig. \ref{fig:trajectory_comparison}. In contrast, the tracing on the stellarator case follows the characteristic geometry of the stellarator \figref{fig:true_stel}.

\subsubsection{Different Trajectories}
In toroidal fusion plasmas like tokamaks, neoclassical transport arises from collisions in the toroidally inhomogeneous magnetic field, which is fundamentally different from classical transport in a uniform field. In a simple cylindrical model (“classical” limit), cross‐field diffusion is set by the random walk of gyrocenters (step size, the Larmor radius) \cite{helander2002collisional,wesson2011tokamaks}. However, the tokamak’s toroidal geometry makes the field stronger on the inboard side, so particles with small parallel velocity become magnetically trapped and bounce between mirror points on the outboard side. These trapped particles execute banana‐shaped orbits in the poloidal plane. \cite{helander2002collisional}. 

By choosing an appropriate initial position and velocity for some particles, we should be able to reproduce banana and trapped orbits in the tokamak. These serve as a first step experiment to check if the stellarator code in JOREK is suitable to reproduce the effects of neoclassical transport and consequently see if the results from the pure tokamak model are in agreement.

\begin{figure}[H]
    \centering
    \hspace*{-4cm}  
    \includegraphics[width=1.2\linewidth]{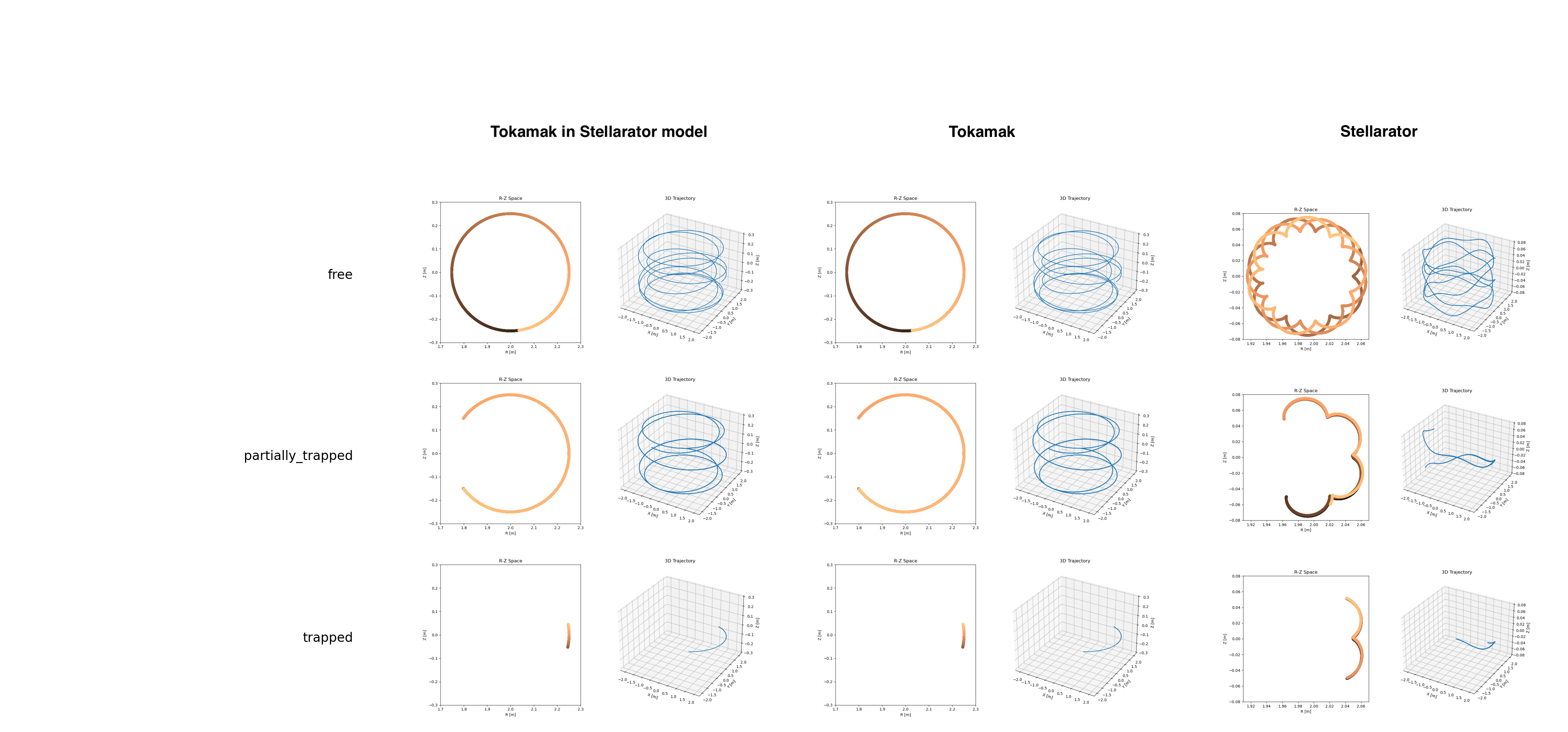}
    \caption{The three types of particle trajectories in the tested cases}
    \label{fig:three_cases}
\end{figure}

As expected, the results show perfectly matching results between the models, assuring compatibility for particle pushing, see Fig. \ref{fig:three_cases}.

\subsubsection{Particle Projection}

While the pusher allows for particles to follow the field calculated on the fluid timestepping, at some point, it is necessary to project the particles on the fluid mesh to retrieve useful properties on the particle moving through the plasma. To achieve this, in the axis-symmetric tokamak, each harmonic is retrieved by solving a separate linear system. 

As this is the complementary step for kinetic simulations, more tests have been performed to ensure agreement in the different models. To achieve this, a special three-dimensional Gaussian particle distribution has been employed spanning across the two poloidal dimensions and on the toroidal dimension.

\begin{figure}[H]
    \centering
    \includegraphics[width=.85\linewidth]{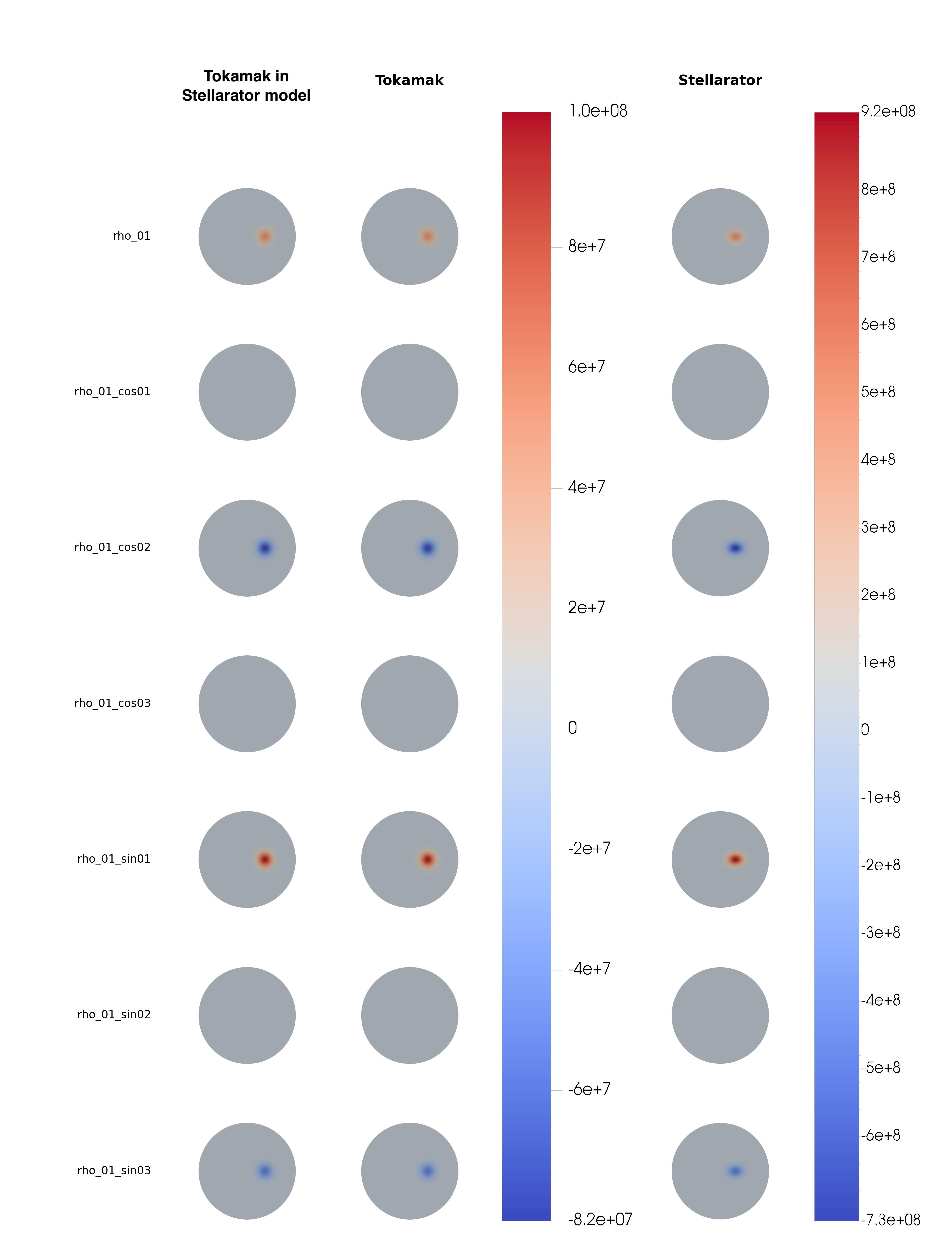}
    \caption{Projection of three-dimensional Gaussian distribution}
    \label{figure:uncoupled_gaussian}
\end{figure}

The result of the aforementioned distribution when projected is a two-dimensional Gaussian, decomposed in its harmonics \figref{figure:uncoupled_gaussian}.

As expected, the results in the two tokamak-like cases are sound across models. Here tough we can have a first appreciation of the problem revolving around projection in the stellarator case when using the uncoupled harmonics system, since the geometry is non-axisymmetrical we cannot treat each harmonic separately and solving their respective linear system, we have to couple all the harmonics together and account for the three dimensional variation of the grid along the toroidal direction in the Jacobian.

\subsection{Implementation of the coupled harmonics}

The mathematical derivation done in \eqref{eq:coupled_harmonics} highlights the system for which we are trying to solve; this required some significant modifications to the code, as the infrastructure heavily relied on splitting the system among the harmonics. 

\subsubsection{Changes in the matrix system}

Before jumping directly to the results, it is important to verify that our mathematical assumptions and conclusions are well reflected in the practical solution implemented.

The first concrete difference we should be able to see is a Jacobian changing along the toroidal angle:

\begin{figure}[H]
    \centering
    \includegraphics[width=1\linewidth]{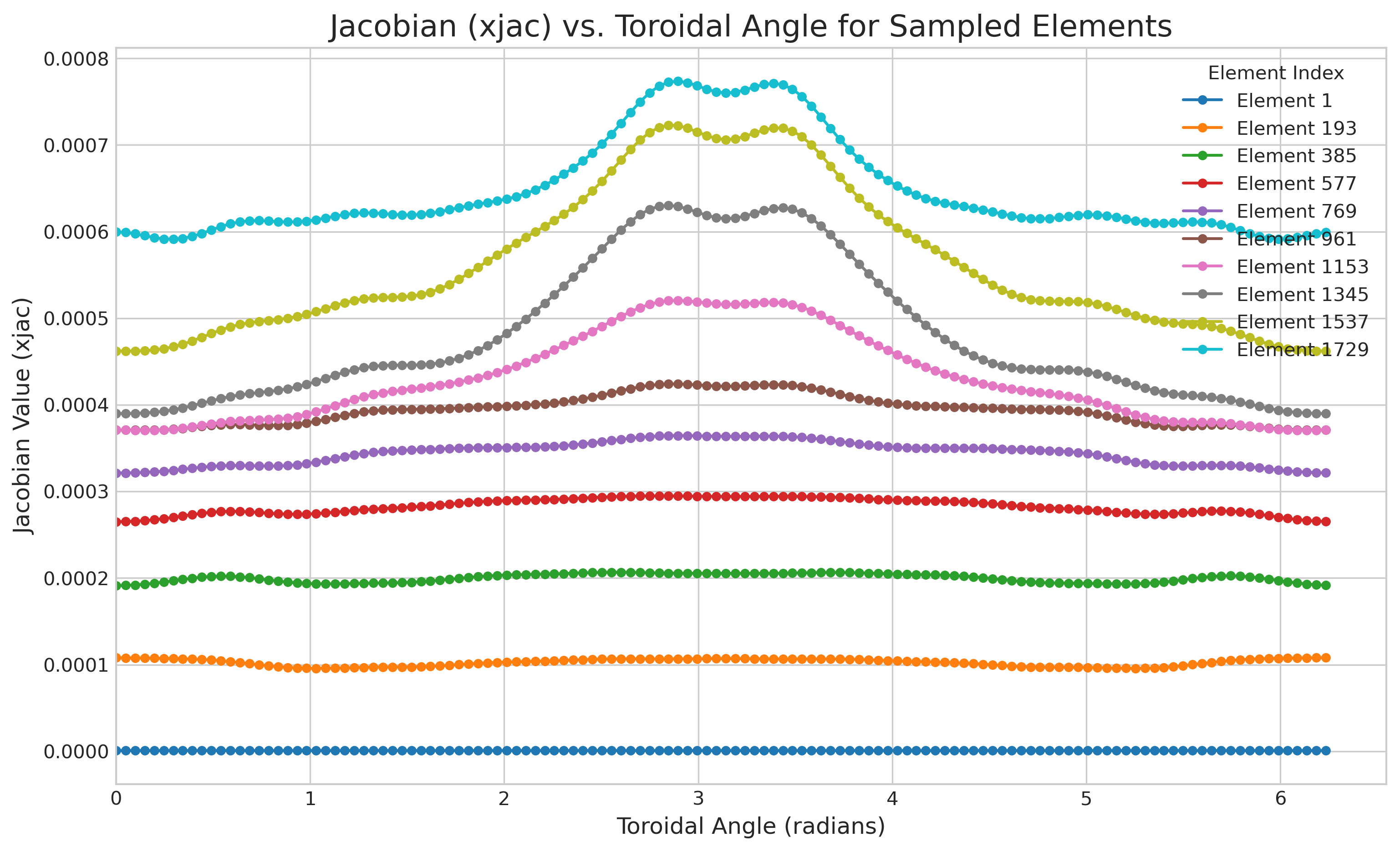}
    \caption{Dependence of the Jacobian on the toroidal angle for several mesh elements. The plot shows a stronger variation for the innermost elements due to their higher geometric distortion near the mesh axis.}
    \label{fig:xjac}
\end{figure}

As evident from Fig. \ref{fig:xjac}, the Jacobian now presents a dependence on the toroidal angle. Interestingly, we can also observe a stronger influence on the Jacobian in the innermost elements. This is due to the higher distortions towards the mesh axis of the elements.

This dependence on the three-dimensional Jacobian also allows us to create a meaningful system to solve, considering all the harmonics, which is immediately evident from Fig. \ref{fig:matrix_comparison}:

\begin{figure}[H]
    \centering
    \begin{subfigure}[t]{0.48\textwidth}
        \centering
        \includegraphics[width=\linewidth]{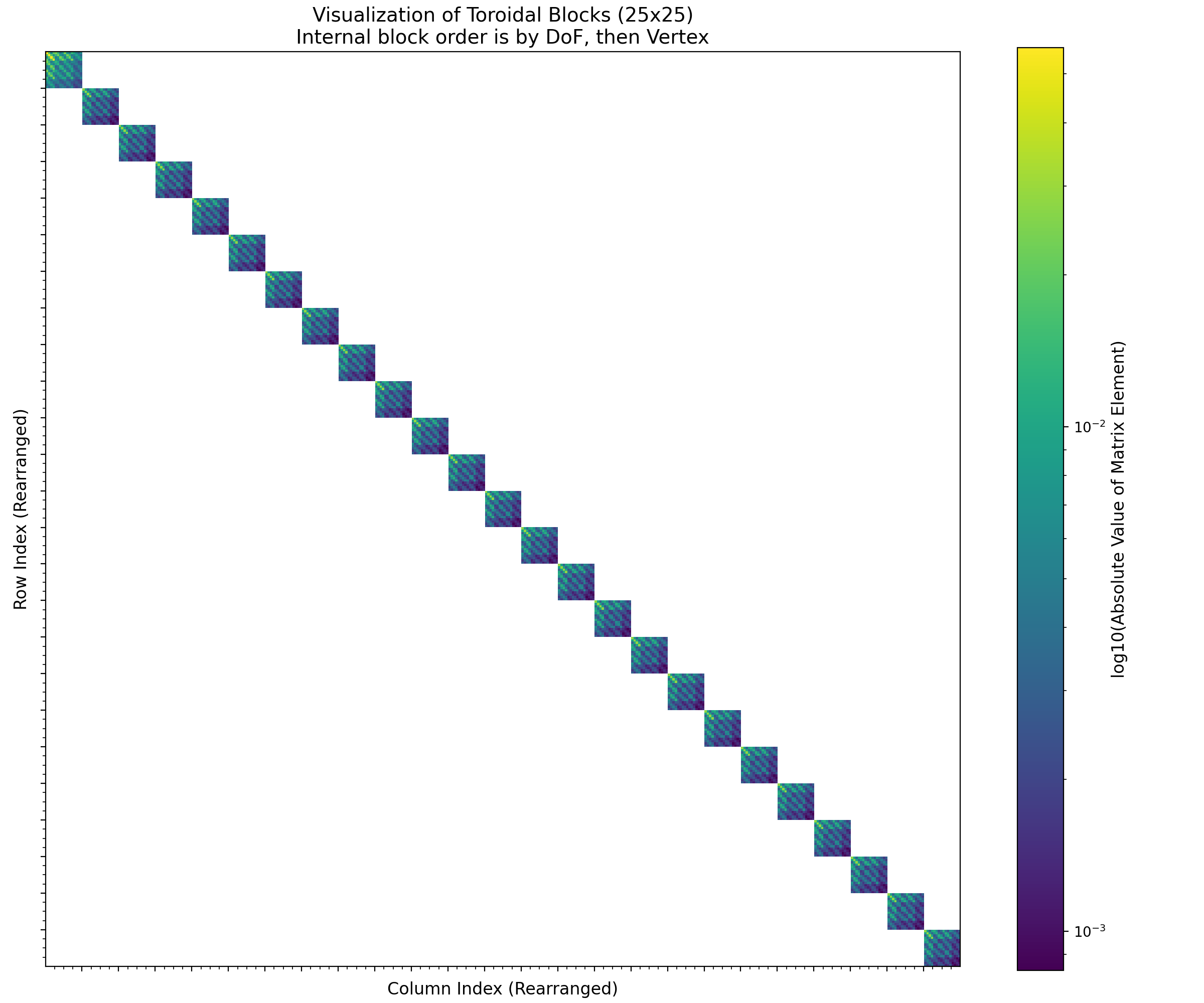}
        \caption{Sparsity pattern of a system matrix where the toroidal modes are uncoupled. Each colored block on the diagonal represents the physics within a single toroidal mode number, $n$. The zero-valued off-diagonal blocks indicate no coupling between different modes.}
        \label{fig:uncoupled_matrix}
    \end{subfigure}
    \hfill 
    \begin{subfigure}[t]{0.48\textwidth}
        \centering
        \includegraphics[width=\linewidth]{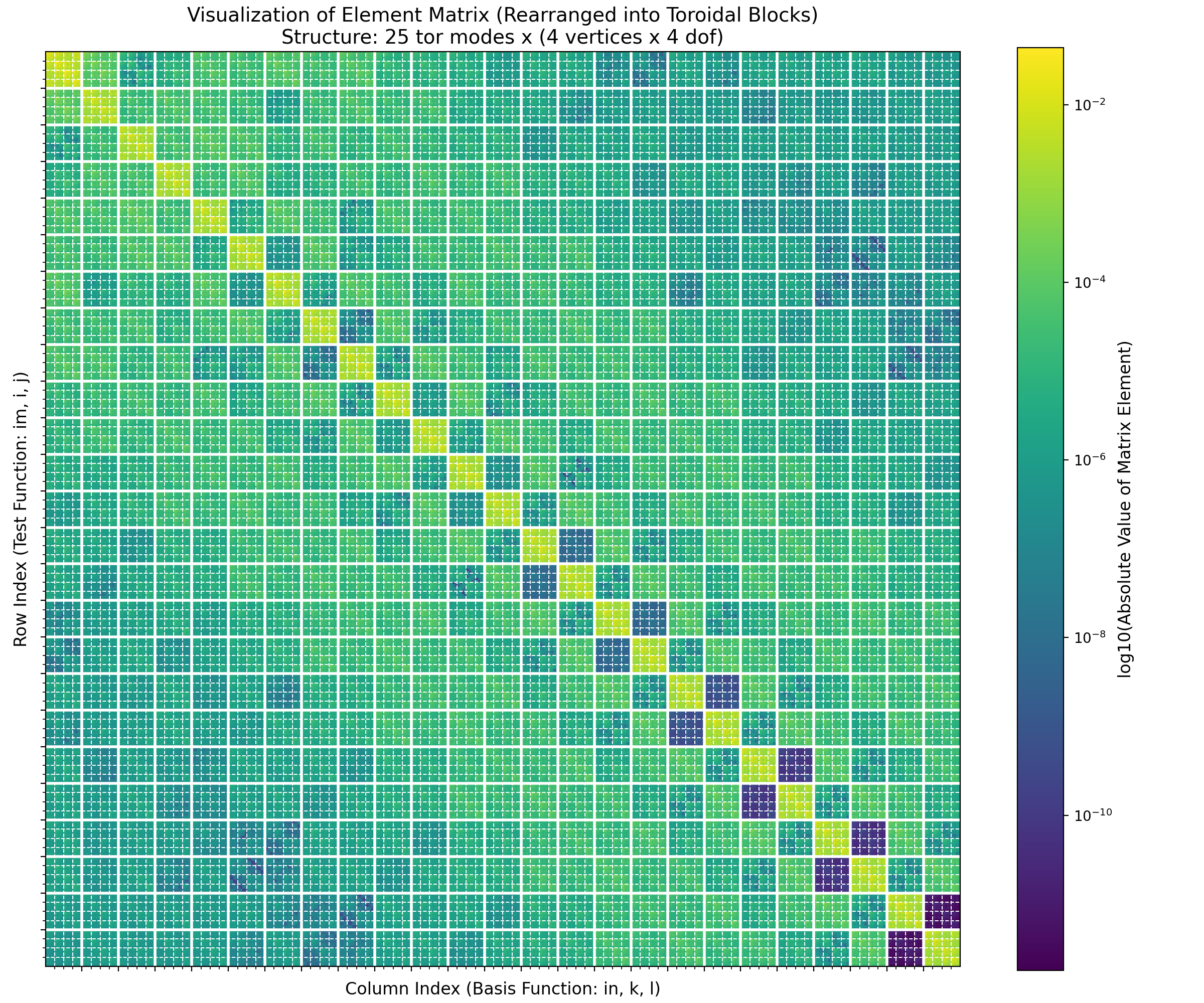}
        \caption{Sparsity pattern of a system matrix with coupling between toroidal modes. The non-zero off-diagonal blocks represent the interaction between different modes (e.g., coupling between modes $n$ and $n\pm1$)}
        \label{fig:coupled_matrix}
    \end{subfigure}
    \caption{Comparison of system matrix structures for a physical model with (a) uncoupled and (b) coupled toroidal modes. The uncoupled case results in a simple block-diagonal matrix, while coupling introduces off-diagonal terms.}
    \label{fig:matrix_comparison}
\end{figure}

\subsection{Coupled projection}

With all the modifications in place, we can now have a qualitative appreciation of the coupled harmonics by revising the projection proposed in Fig. \ref{figure:uncoupled_gaussian} in the coupled scenario.

Before being able to appropriately represent our complex geometry, it was necessary to take into account the non-axysimmetric geometry of the stellarator. As previously discussed, the assumption on the geometry influenced not only the underlying math of the system but also several support diagnostics along with the code, in particular, the vtk export resulting from projecting particles used the 2D basis to represent the poloidal section without taking into account the toroidal angle; for this reason, an extension to 3D geometries has been developed. 

Although the geometry is represented continuously over the toroidal angle $\phi \in [0,2\pi]$, for clarity, this work will primarily show results at a few discrete toroidal planes

\subsubsection{Gaussian distribution on the W7-A}

The first test case considered is the Wendelstein 7-A (W7-A)\cite{w7a}, a classical stellarator. While its geometry is non-axisymmetric, it is not subject to drastic deformation. The three-dimensional variation provides a suitable benchmark to initially validate the coupled harmonic projection scheme.

For these tests, we employ a three-dimensional Gaussian of the form:
\begin{equation}
f(R, Z, \phi) =\exp \left( - \left( \frac{(R - R_0)^2}{2\sigma_R^2} + \frac{(Z - Z_0)^2}{2\sigma_Z^2} + \frac{(\phi - \phi_0)^2}{2\sigma_\phi^2} \right) \right) 
\end{equation}

For the Gaussian projected onto W7-A, the parameters chosen are:
\begin{gather*}
R_0 = 2.02, \quad Z_0 = 0, \quad \phi_0 = \frac{\pi}{2} \\
\sigma_R = \sigma_Z = 0.01,\quad \sigma_\phi = \frac{\pi}{2}
\end{gather*}

\begin{figure}[H]
    \centering
    \begin{subfigure}{0.25\textwidth}
        \includegraphics[width=\linewidth]{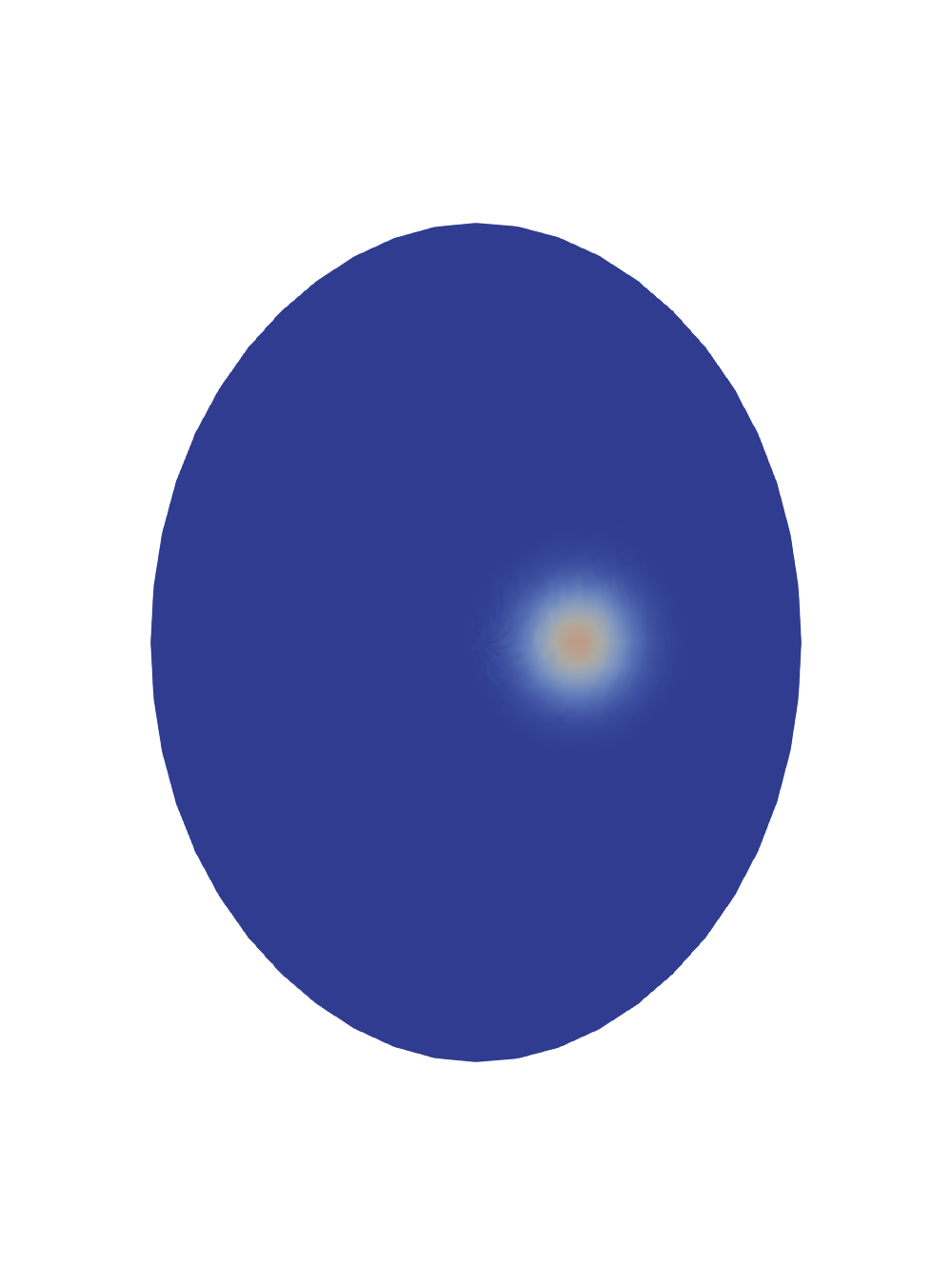}
        \caption{Cross-section at $\phi = 0$.}
    \end{subfigure}
    \hfill 
    \begin{subfigure}{0.27\textwidth}
        \includegraphics[width=\linewidth]{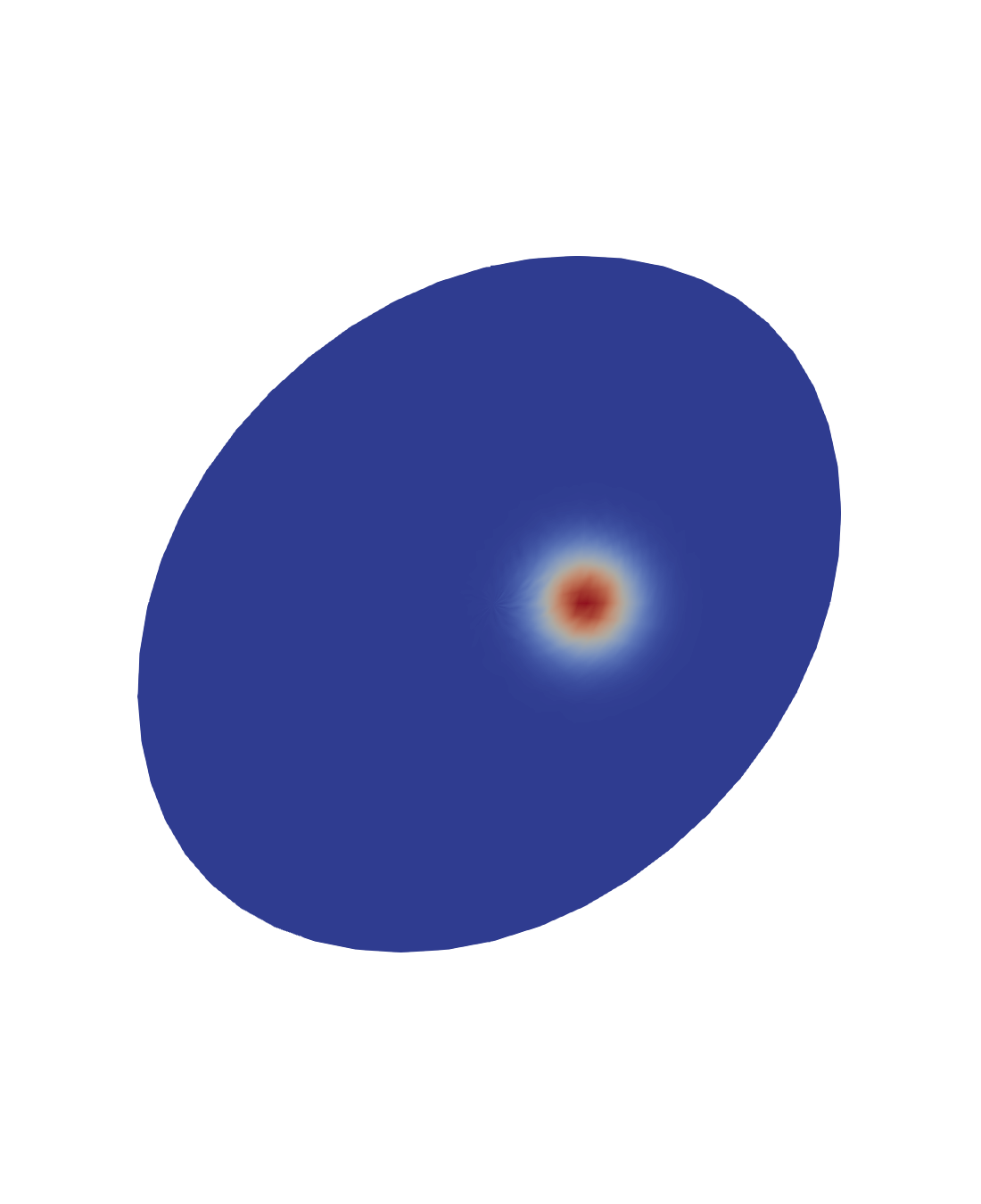}
        \caption{Cross-section at $\phi = \frac{\pi}{2}$.}
    \end{subfigure}
    \hfill 
    \begin{subfigure}{0.29\textwidth}
        \includegraphics[width=\linewidth]{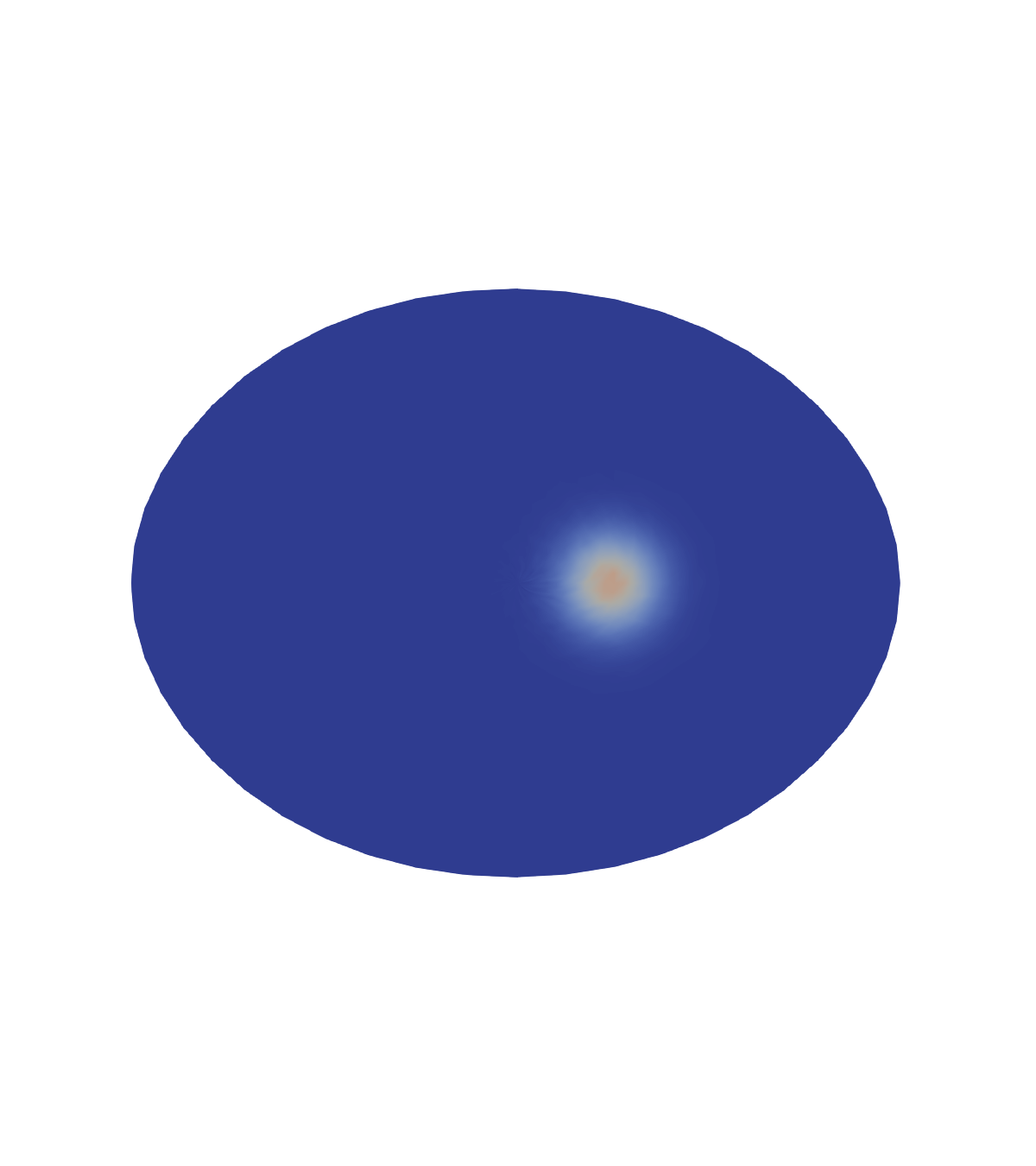}
        \caption{Cross-section at $\phi = \pi$.}
    \end{subfigure}
    \hfill 
    \begin{subfigure}{0.1\textwidth}
        \centering
        \includegraphics[width=\linewidth]{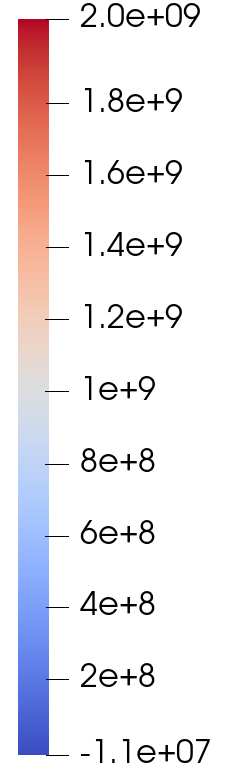}
    \end{subfigure}
    \caption{Visualization of the W7-A stellarator's non-axisymmetric geometry. The poloidal cross-section is shown at three different toroidal angles.}
    \label{figure:w7a}
\end{figure}

As shown in Fig. \ref{figure:w7a}, even for relatively simple non-axisymmetric geometries, the toroidal dependency of the mesh already has a noticeable influence on the solution; taking into account the full Jacobian, such deformation is correctly accounted for and the shape of the Gaussian is recovered correctly.

\subsubsection{Gaussian distribution on the W7-X}

While W7-A provides a suitable starting point, it is crucial to move toward more realistic and contemporary geometries, where a strong dependence on the toroidal direction greatly deforms the mesh shape. For these reasons, further tests have been conducted on the geometry of the W7-X \cite{w7x}, the latest stellarator reactor located at the Max Planck Institute for Plasma Physics in Greifswald. 

W7-X is a quasi-isodynamic stellarator, and the geometry of the reactor closely resembles the geometries of SQuIDs \cite{squids}, a promising future reactor design for which, in recent years, great interest has developed.

Due to the highly shaped mesh, the Gaussian distribution cannot be initialized in the same $RZ$ position like in the W7-A case; for this reason, suitable positions are chosen to showcase the most deformed geometries. 

\begin{figure}[H]
    \centering
    \begin{subfigure}{0.35\textwidth}
        \includegraphics[width=\linewidth]{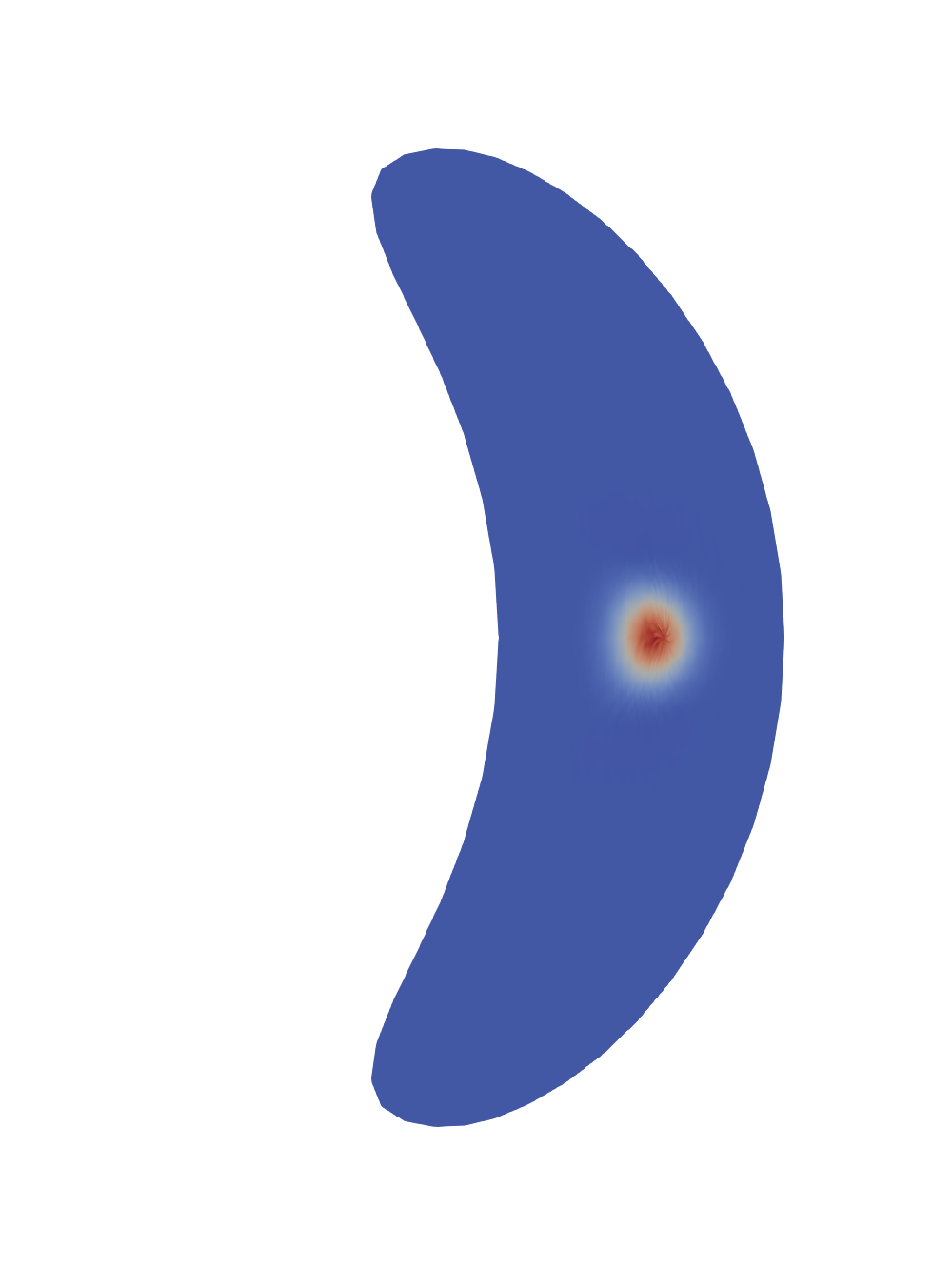}
        \caption{Cross-section at $\phi = 0$ with $R_0 = 6$.}
        \label{fig:w7x_0}
    \end{subfigure}
    \hfill 
    \begin{subfigure}{0.48\textwidth}
        \includegraphics[width=\linewidth]{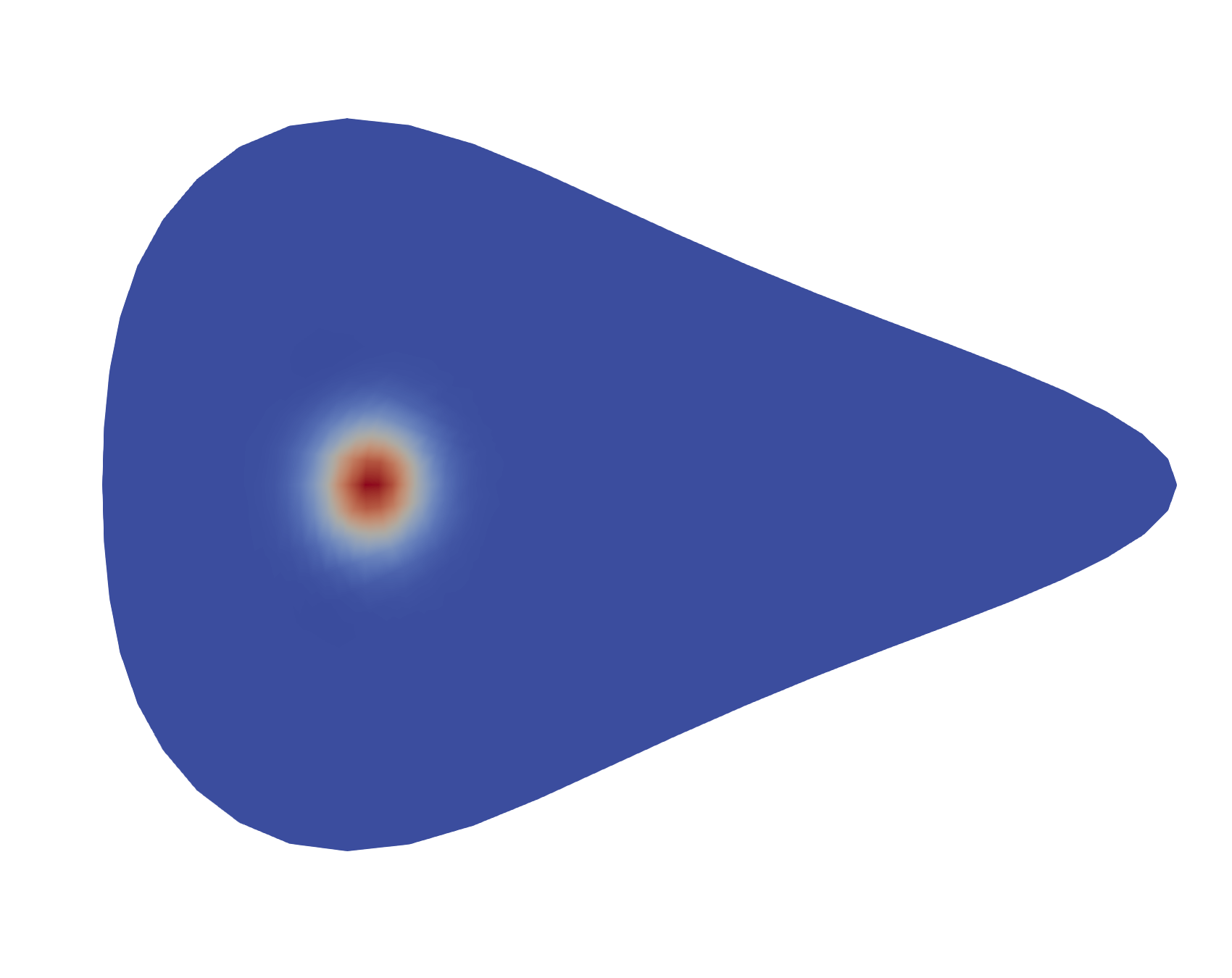}
        \caption{Cross-section at $\phi = \pi$ with $R_0 = 5$.}
        \label{fig:w7x_pi}
    \end{subfigure}
    \hfill 
    \begin{subfigure}{0.1\textwidth}
        \includegraphics[width=\linewidth]{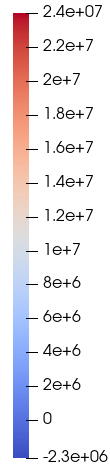}
    \end{subfigure}
    \caption{Projection of a Gaussian distribution onto the W7-X geometry. The significant deformation of the poloidal cross-section between the toroidal angles $\phi=0$ and $\phi=\pi$ showcases the capability of the coupled projection scheme on a highly shaped, realistic geometry.}
    \label{figure:w7x_projection}
\end{figure}

The Results shown in Fig. \ref{figure:w7x_projection} use a 3D Gaussian with parameters: $Z_0 = 0, \sigma_R = \sigma_Z = 0.45,\quad \sigma_\phi = \frac{\pi}{16}$

\subsection{Convergence Tests}

While the qualitative analysis shows promising results, it is vital to present quantitative studies as evidence of the scheme's accuracy. For this, a range of convergence tests has been performed by projecting some known analytical functions onto the W7-X grid. The idea is to measure the change of error as we increase the number of toroidal harmonics used to compute the solution. With the uncoupled projection, no convergence is expected since the geometrical coupling of harmonics is neglected. In contrast, for the coupled scheme, we expect an exponential, spectral \cite{Spectral_Methods}, convergence.

We choose the $L^2$ norm between the original analytical function and the numerical solution from the projection on the FEM grid. By increasing the resolution of the solution, through the increase in the number of toroidal harmonics, we expect this error to decrease faster than the uncoupled counterpart. These tests have been performed on three different functions and are reported below:

\begin{figure}[H]
    \centering
    \includegraphics[width=0.8\linewidth]{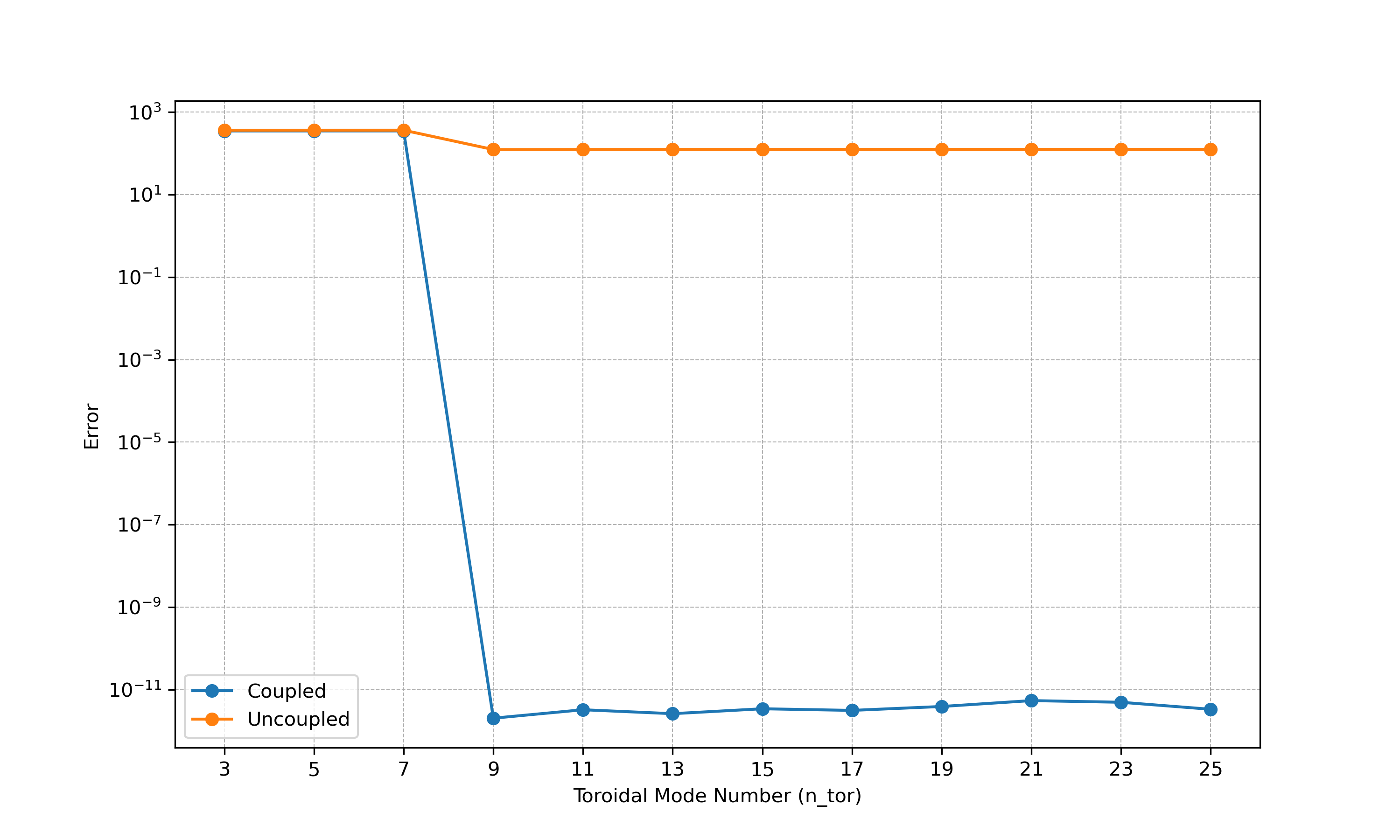}
    \caption{Performance of the coupled vs. uncoupled schemes for the function $\cos^2(2\phi)$. The uncoupled system fails to converge, while the coupled system shows a sharp convergence. The error decays precisely at $n_{\text{tor}}=9$, which corresponds to a maximum mode number of $M=4$ in our basis set ($n_{\text{tor}} = 2M+1$), capturing the dominant harmonic in the function's analytical form. The remaining error at higher resolution is close to the machine epsilon for double-precision floating points. }
    \label{fig:conv_cos_square} 
\end{figure}

The first function under consideration is a $cos^2(2\phi)$, whose Fourier expansion has a periodicity of 4 ($\frac{1}{2}+\frac{1}{2}cos(4x\phi)$). The behavior is perfectly in line with the theoretical assumptions: the sharp increase in accuracy is explained by the introduction of the harmonic with the proper periodicity into the spectral decomposition \figref{fig:conv_cos_square}. This pathological case highlights how the uncoupled system is fundamentally unable to approximate the function to a meaningful level of accuracy.

\begin{figure}[!h]
    \centering
    \begin{subfigure}[t]{0.49\linewidth}
        \includegraphics[width=\linewidth]{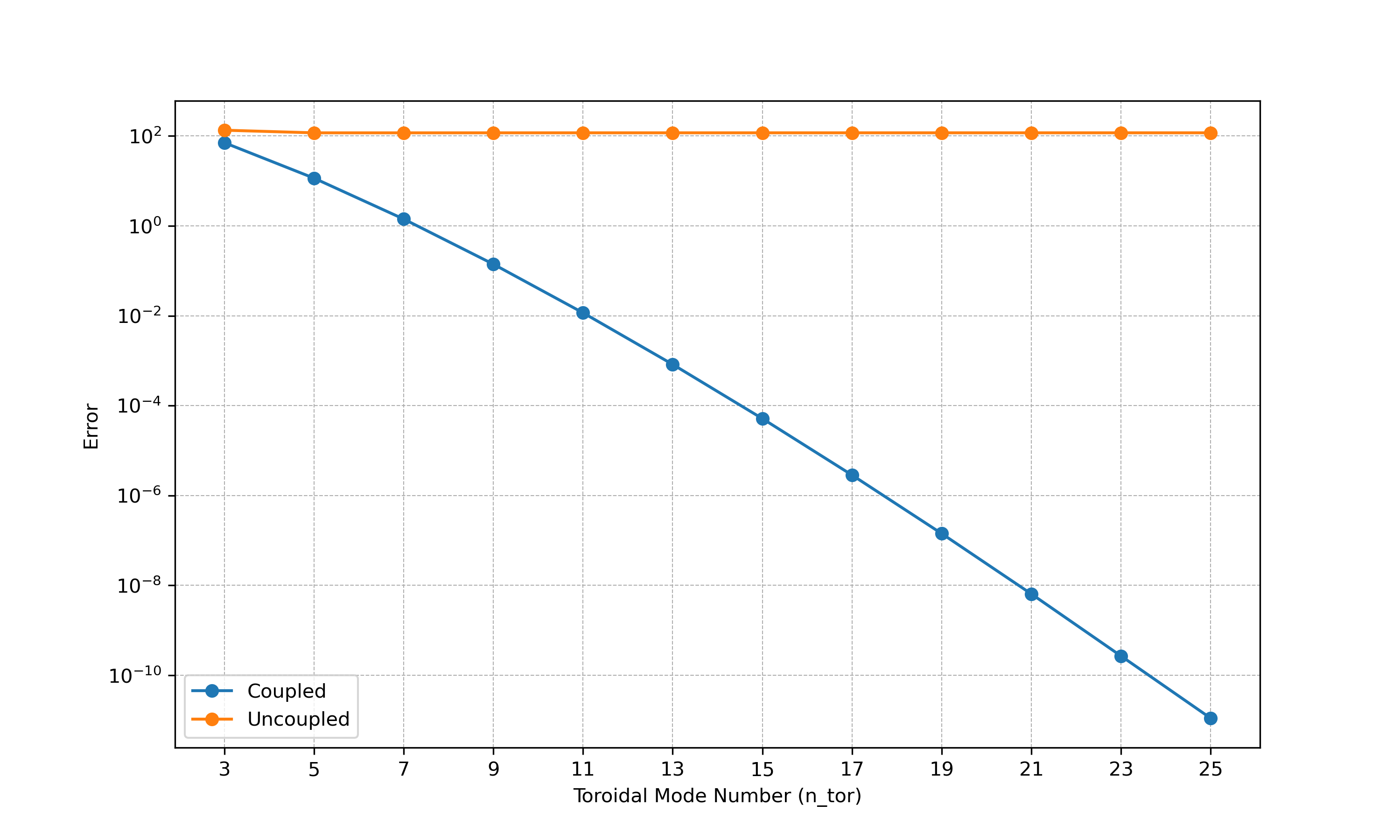}
        \caption{Convergence for a periodized Gaussian function. The function is smooth and periodic.}
        \label{fig:gauss_periodic}
    \end{subfigure}
    \hfill 
    \begin{subfigure}[t]{0.49\linewidth}
        \includegraphics[width=\linewidth]{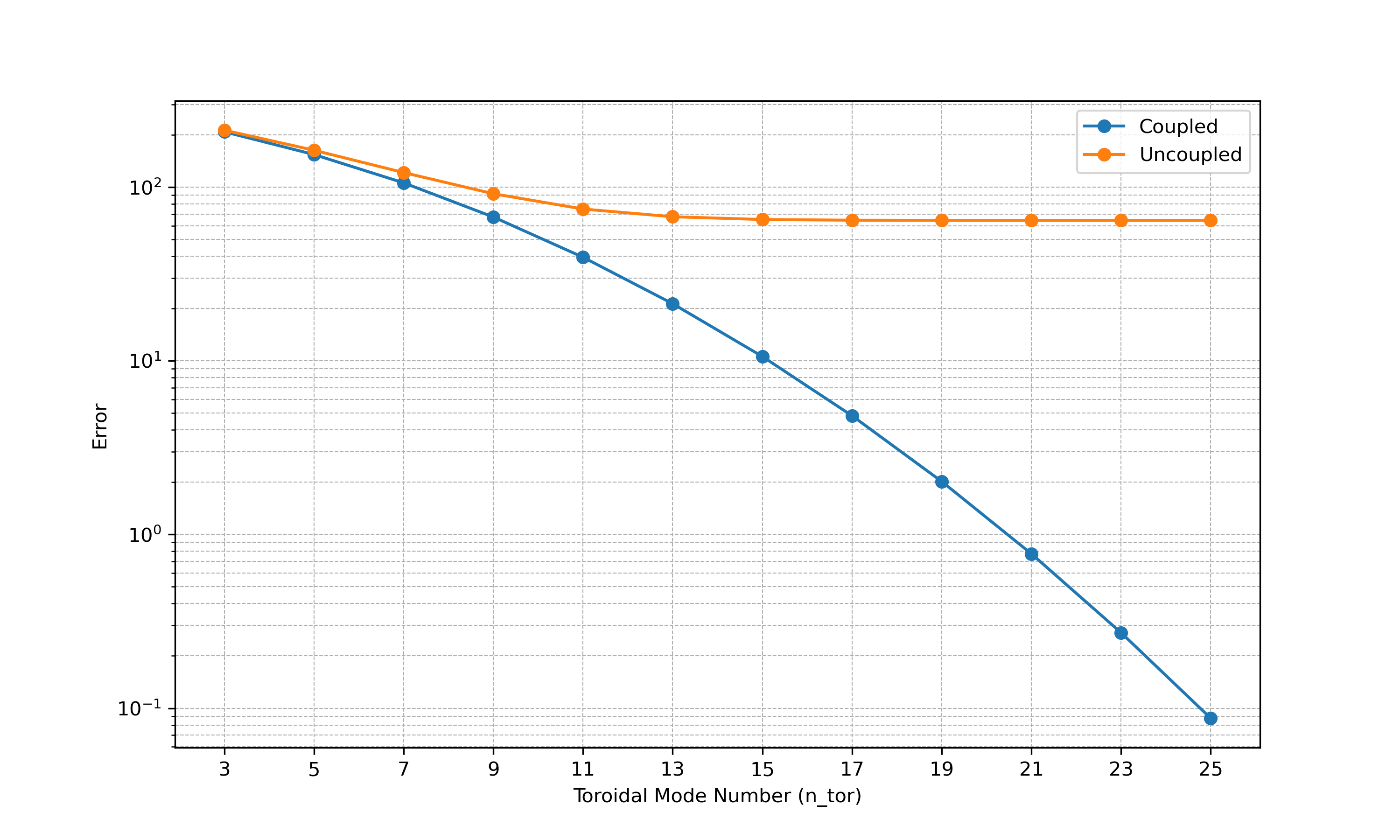}
        \caption{Convergence for a non-periodic Gaussian function. The convergence rate is slower, and the uncoupled approximation is good enough for a small $n_{tor}$ number.}
    \end{subfigure}
    \caption{Convergence tests for two Gaussian functions projected on the W7-X grid.}
    \label{fig:gauss}
\end{figure}

In order to validate the method against more realistic distribution scenarios, tests were performed using Gaussian functions \figref{fig:gauss}. The first scenario presents the perfect candidate for the convergence study. By using a periodic and smooth ($C^\infty$) function such as $e^{cos(\phi)}$, we can appreciate the spectral convergence of the method. The second case, a standard Gaussian on the toroidal direction of equation: $\exp\left( -\frac{(\phi - \pi)^2}{2(0.3)^2} \right)$, demonstrates how the uncoupled system can almost approximate the results from the coupled scenario for a small number of harmonics, usually this is not the case for production cases as the complex physics needs a high number of harmonics to be properly represented.

\subsection{FFT acceleration of the matrix construction}

As described in sec. \ref{sec:fft_math}, the integral on the toroidal direction can be efficiently evaluated by means of the FFT algorithm. This strategy was already employed in the construction of the matrix in the fluid time stepping, but since the toroidal dependency is missing in the tokamak case, the strategy had to be implemented from scratch for the projection system. 

The general strategy implemented was to collect all the toroidally independent parts of the integral together, and then, depending on whether the user wants to run the code with or without the FFT, either add the direct integration contribution or compute the coefficients via FFT. For the FFT computation, we leveraged the FFTW library \cite{fftw}.

The actual system to be solved for is slightly different from the one presented earlier. In order to get a smoother projection of the kinetic quantities, it is sometimes useful to use some filters applied to the derivatives of the solution. For these terms, a separate FFT needs to be performed; thus, two different sets of FFTs are performed separately and later reassembled together in one single matrix per element. 

\begin{figure}[h!]
    \centering
    \includegraphics[width=0.85\linewidth]{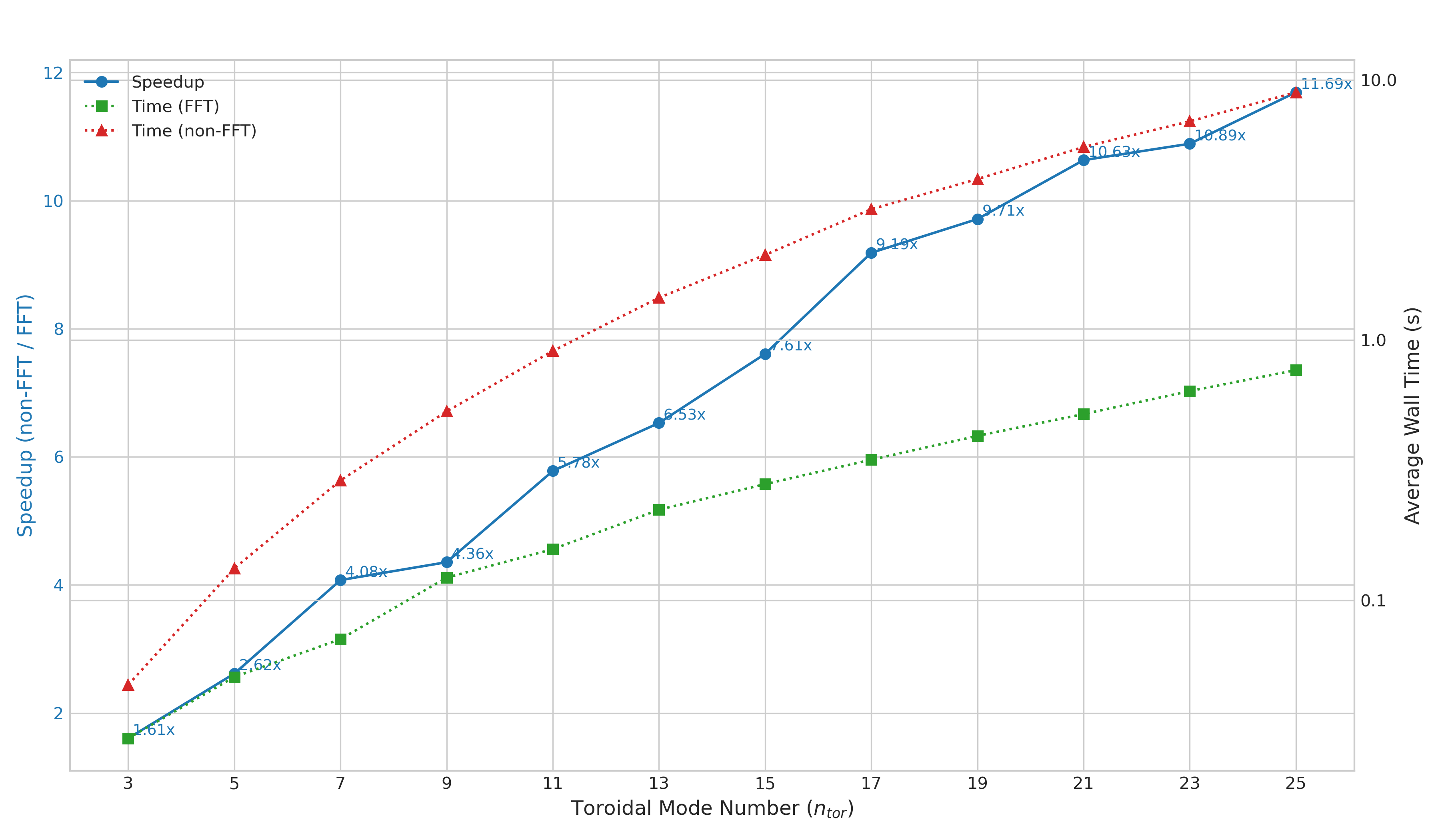}
    \caption{Comparison of wall time for the matrix assembly with (FFT) and without (non-FFT) considering $n_{plane} = 2n_{tor}$ the Fast Fourier Transform acceleration, as a function of the number of toroidal harmonics ($n_{tor}$). The speedup, calculated as the ratio of non-FFT time to FFT time, demonstrates a significant performance gain, aligned with the theoretical linear speedup with respect to $n_{tor}$.}
    \label{fig:fft_speedup}
\end{figure}

An excellent way to demonstrate the effectiveness of this implementation is to compare the performance with and without the FFT acceleration. Fig. \ref{fig:fft_speedup} shows the average wall time required to build the projection matrix for both the direct integration (non-FFT) and the FFT-based methods alongside the speedup, plotted against the number of toroidal harmonics used ($n_{tor}$).

As can be noted in Fig. \ref{fig:fft_speedup}, while the number of toroidal harmonics increases, the computational cost of the direct integration method (represented by red triangles) increases quickly. The wall time for the FFT-based method (green squares), on the other hand, shows a considerably softer increase. Performance is significantly improved as a result of this scaling difference. The blue line on the plot represents the speedup, which is the ratio of the non-FFT time to the FFT time. With $n_{tor}$, it grows almost linearly, and for $n_{tor} = 25$, it achieves a notable speedup. 

\subsection{The RTree}
\label{sec:rtree}

A crucial missing step for a complete working simulation lies in the association of each particle with the corresponding element. This step is needed since each particle lives in the real space $RZ\phi$, but the finite element formulation requires an element-wise location of the particles to project the relevant quantities. 

This problem can be reduced to a classical spatial indexing problem, for which many solutions have been developed over the years, depending on the specific application requirements. 

For the problem at stake, since the mesh naturally partitions the space into regions, a region proposal algorithm is required. For these reasons, the RTree \cite{RTree} algorithm and data structure were the most natural choice and are already used in the 2D tokamak applications of the code.

The underlying idea being that the bounding boxes in three dimensions of each element must cover the entire element, this is achieved by finding the minimum and maximum points of the element along two different poloidal section located at $\phi_1, \phi_2$ and defining the bounding box as $([R_{min},R_{max}], [Z_{min}, Z_{max}], [\phi_1, \phi_2])$.

To find the 2D minimum bounding box at a given poloidal section, we can look for the points at which the derivative is zero along each of the four element sides, as briefly explained in sec. \ref{sec:AABB_bezier}.

\begin{figure}[H]
    \centering
    \begin{subfigure}[t]{0.235\textwidth}
    \centering
        \includegraphics[width=\linewidth]{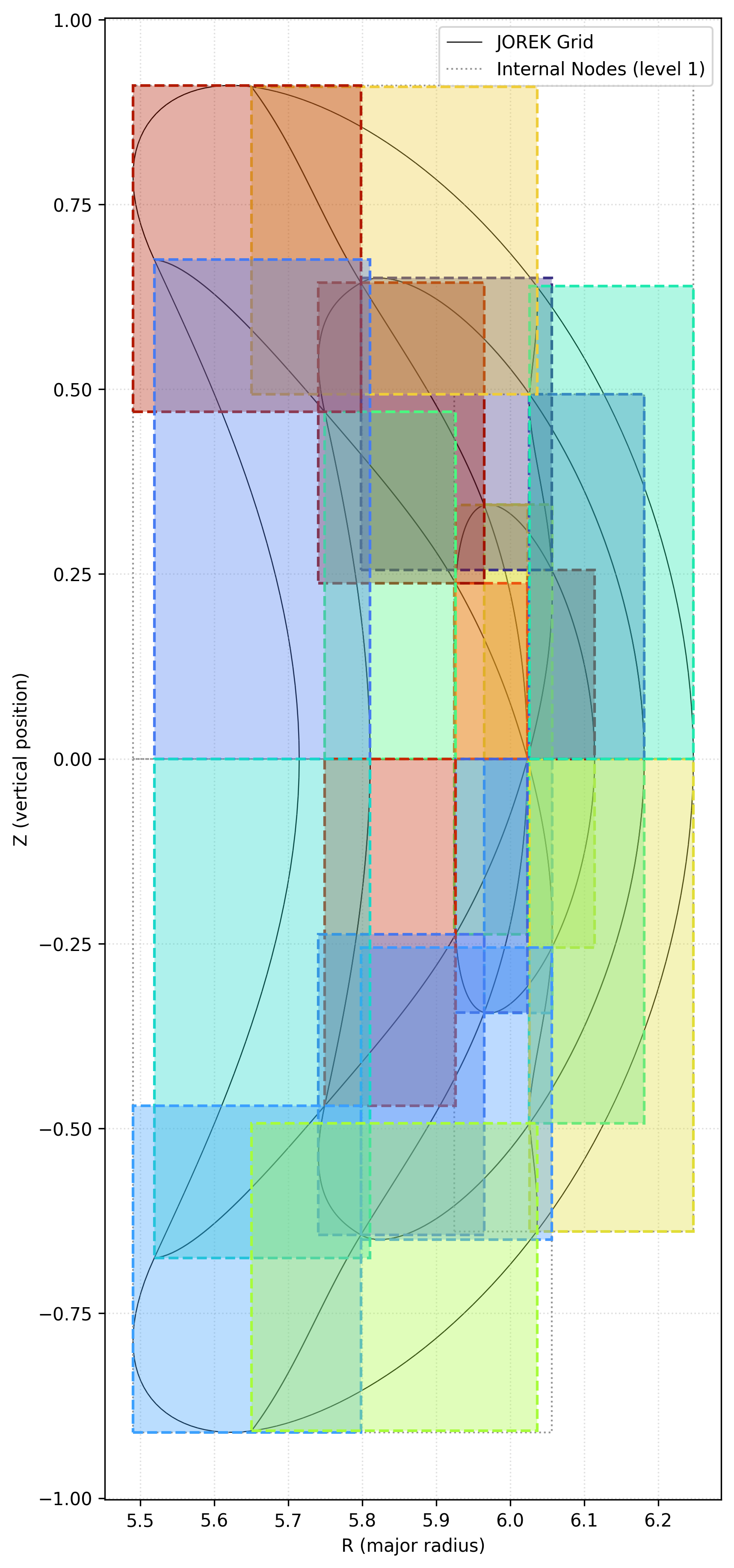}
        \caption{R-Tree structure for the poloidal plane $\phi = 0$}
        \label{fig:rtree_slice_zero}
    \end{subfigure}
    \hfill
    \begin{subfigure}[t]{0.74\textwidth}
    \centering
        \includegraphics[width=\linewidth]{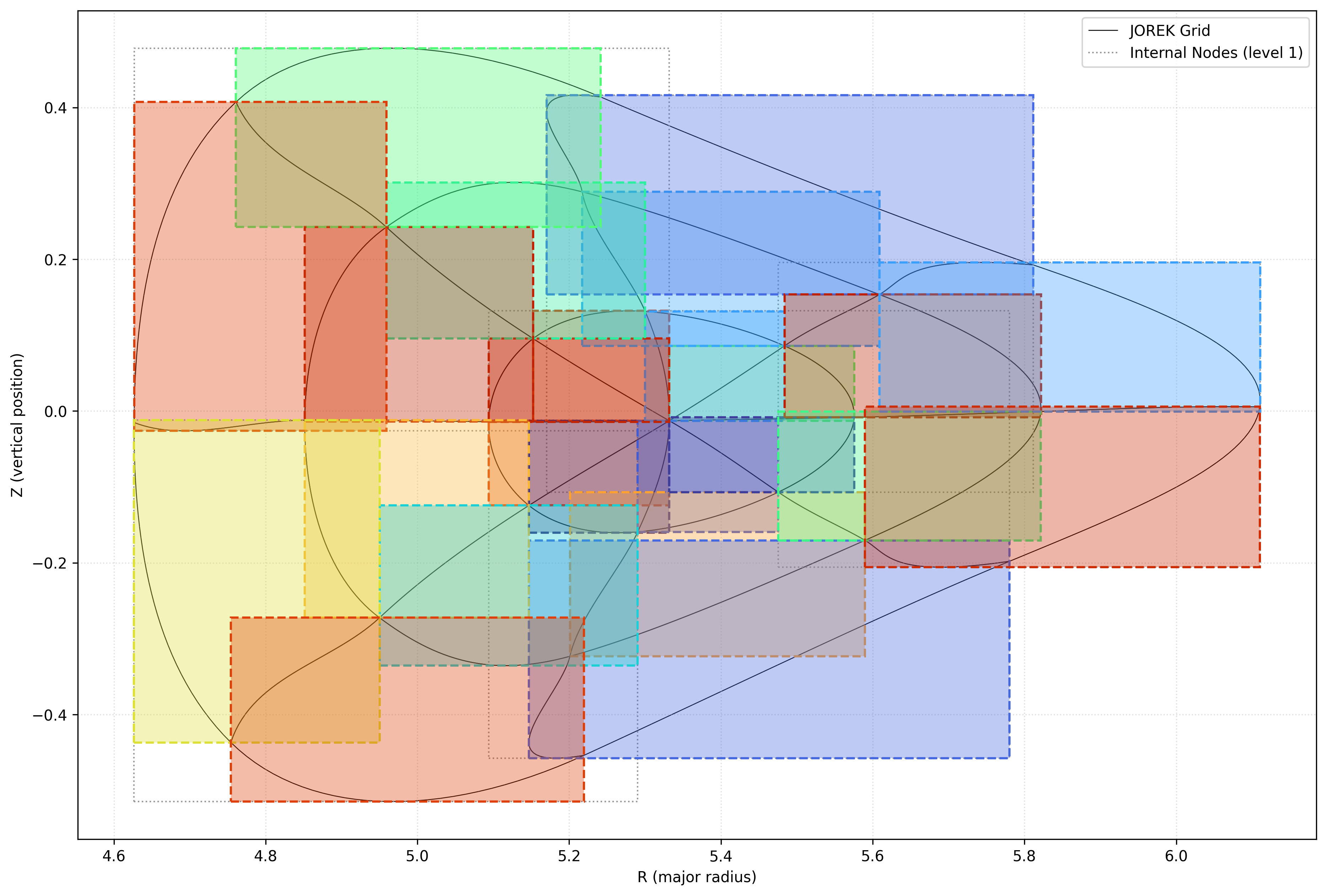}
        \caption{R-Tree structure for the poloidal plane $\phi = \pi$}
        \label{fig:rtree_slice_phi}
    \end{subfigure}
    \caption{Visualization of the 2D R-Tree spatial indexes for the W7-X grid at two different toroidal angles. Each rectangle represents the 2D bounding box of a single mesh element at that specific poloidal cross-section. The significant change in the bounding box layout between (a) and (b) illustrates the strong geometric variation along the toroidal direction.}
    \label{fig:rtree2D_visualization}
\end{figure}

\begin{figure}[H]
    \centering
    \includegraphics[width=.98\linewidth]{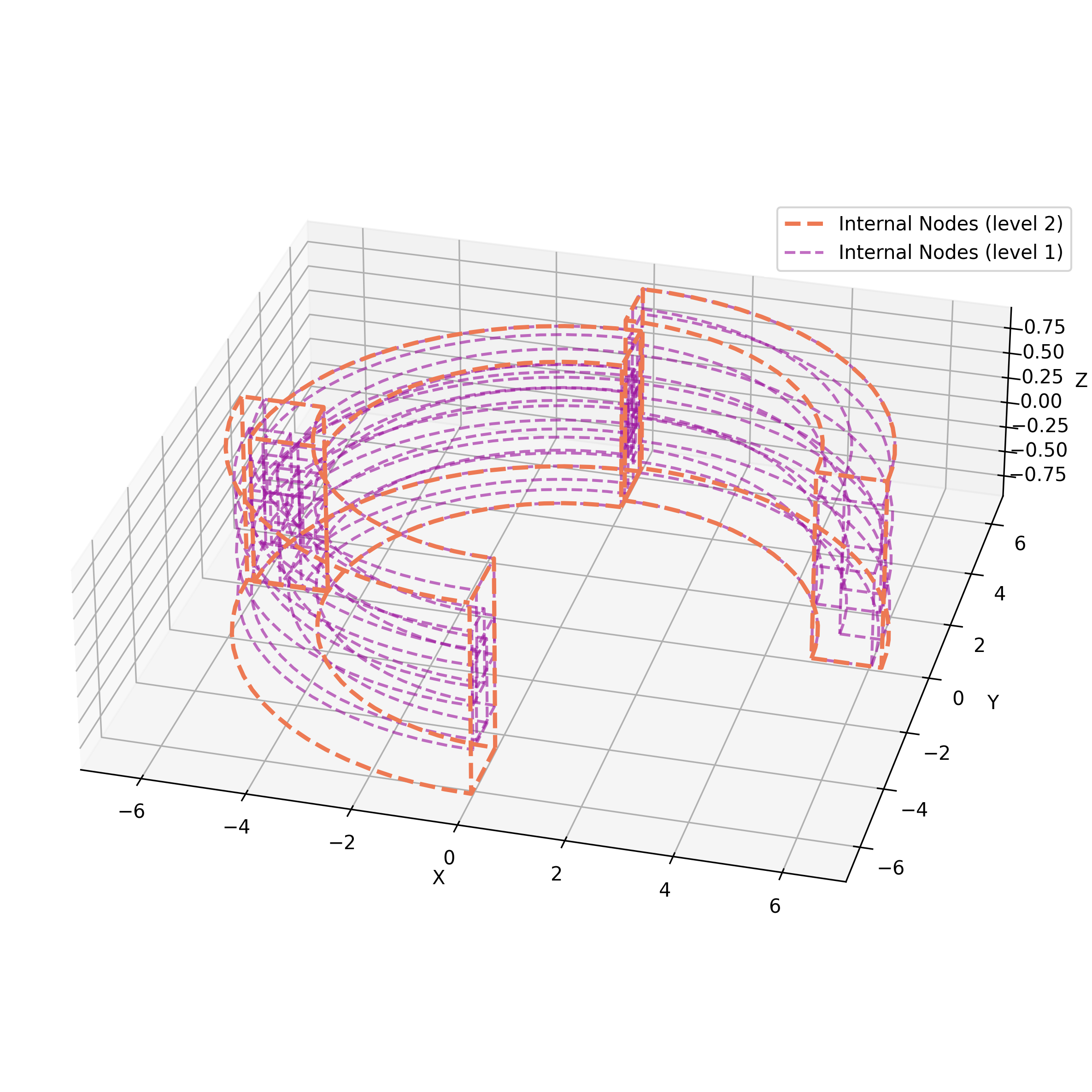}
    \caption{The 3D R-Tree data structure used for efficient particle lookup in the W7-X stellarator geometry. The image shows the 3D bounding boxes that form the tree's nodes, creating a hierarchical map of the simulation domain. This spatial partitioning is a critical component for associating particles with their host finite elements in the non-axisymmetric grid.}
    \label{fig:rtree3D}
\end{figure}

With the minimum of the element in the poloidal section \figref{fig:rtree2D_visualization}, we chose to consider the width of the bounding box in the toroidal direction as the distance between two different poloidal sections, thus creating our three-dimensional bounding box \figref{fig:rtree3D}.

The introduction of the third dimension on both the RTree algorithm and the local coordinate search introduces a significant overhead. This overhead should still be bounded by a factor proportional to the number of particles, for the association step between the particles and elements, and a constant, for the element creation. 

\begin{figure}[h]
    \centering
    \includegraphics[width=1\linewidth]{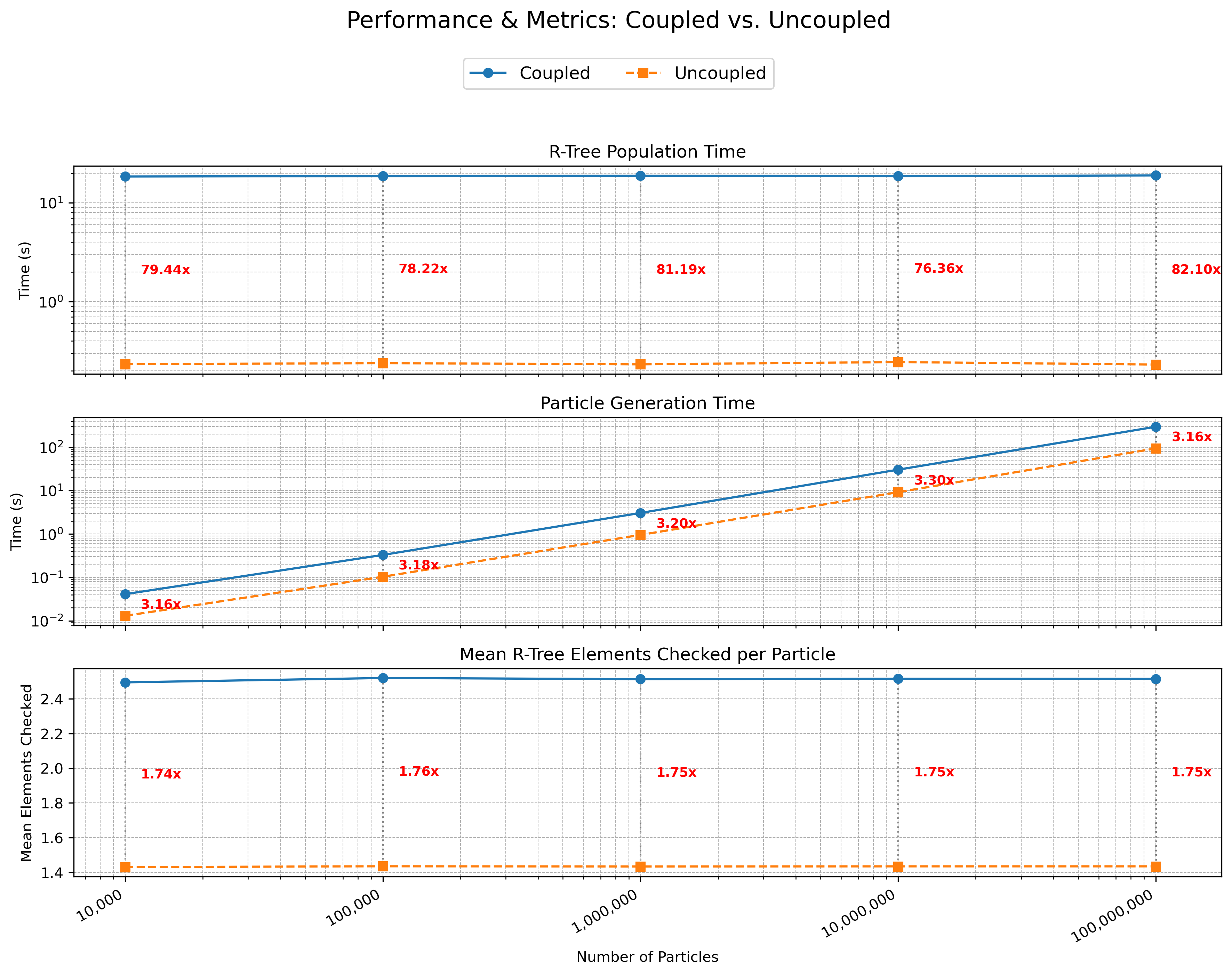}
    \caption{Performance comparison of the R-Tree implementation for element creation and particle association. The plot highlights the computational overhead introduced by the third dimension.}
    \label{fig:rtree_perf_comparison}
\end{figure}

As we can see in Fig. \ref{fig:rtree_perf_comparison}, we have a worst-case scenario of 80x speed-down for the tree creation. This is a favorable result, as a naive slicing of the toroidal direction into $n_{planes} = 128$ might suggest a 128x slowdown. The improved performance is due to the R-Tree's hierarchical structure, which efficiently groups elements and prunes the search space, avoiding a linear scaling with the number of poloidal planes.

As per the particles association itself, adding a third dimension was bound to be computationally expensive; on average, we measure a 3x speed down, for which a small amount is attributable to the imperfect discretizations of the toroidal direction requiring, on average, 1.65x more elements to lookup.

\section{Conclusion}
\label{sec:conclusion}
The development of strong, predictive simulation capabilities is essential to the search for commercially viable fusion energy, particularly for the geometrically challenging stellarator concept. A fundamental challenge in this field is the precise application of a hybrid fluid-kinetic model in a non-axisymmetric framework. Decoupled projection schemes used in simpler tokamak simulations are physically incorrect for stellarators because the toroidal Fourier harmonics are coupled by the intrinsic three-dimensional nature of the magnetic field.

The main contribution of this work is the design, implementation, and validation of a novel, globally coupled particle-in-cell projection scheme within the JOREK finite element code. This technique addresses the core challenge by building and solving a single, cohesive linear system over the entire 3D, non-axisymmetric mesh. By self-consistently accounting for the geometric coupling between all toroidal harmonics, the system ensures a high-fidelity transfer of information from the kinetic particle population to the fluid grid. An effective Fast Fourier Transform (FFT) based approach was used to make this computationally demanding global matrix construction possible. Qualitative projections onto realistic Wendelstein 7-X geometries demonstrated the scheme's ability to capture complex shaping. More significantly, quantitative convergence tests conclusively demonstrated the method's superiority: the coupled scheme achieved the theoretically expected spectral convergence for smooth analytical functions, whereas the uncoupled approach failed to converge, highlighting the necessity of this global treatment.

As a second significant contribution, an effective particle localization algorithm was implemented to address the prohibitive computational cost of a global scheme. We developed a scalable solution for associating millions of particles with their host finite elements by creating a three-dimensional R-Tree spatial index specific to the JOREK mesh structure. Performance analysis verified that this algorithm tackles the 3D search's complexity with a reasonable overhead, keeping the hybrid model tractable for large-scale, high-performance computing.

In addition to providing a solid and verified framework, this work opens up a number of fascinating directions for further investigation.  In order to study how kinetic effects drive or dampen MHD instabilities, the most immediate next step is to enable the feedback of the kinetic particles on the fluid.  A distributed memory model with MPI could be used to parallelize the assembly of the global matrix and right-hand-side vector, further improving scalability. 

Although they come with some trade-offs, other possible improvements have been taken into consideration. One of them might be to find the true extrema of each element's geometry; for example, a Newton iteration could be used to determine the bounding boxes of the R-Tree with greater precision; however, this is probably too computationally costly for a marginal gain. Likewise, iterative solvers \cite{iterativeSolvers} are suboptimal to solve the projection system, even though they are frequently used for large linear systems. The use of direct solver libraries \cite{MUMPS:1, pastix, strumpack} based on a one-time matrix factorization is much more efficient because the projection system is solved repeatedly with the same matrix but a changing right-hand side.

Together, these contributions eliminate a significant obstacle to running high-fidelity, non-linear hybrid simulations of stellarator plasmas. This validated framework makes a previously unattainable range of predictive studies possible, enabling the community to better understand, optimize, and design the next generation of fusion energy devices and advancing our progress toward clean, sustainable energy.
\bibliography{bibliography.bib}



\cleardoublepage

\section*{Algoritmi e ottimizzazioni per la simulazione globale e non-lineare di stellarator tramite modelli ibridi fluido-cinetici con il metodo degli elementi finiti}

Il progresso della ricerca sull’energia da fusione nucleare richiede lo sviluppo di modelli predittivi per i plasmi nei reattori stellarator. Tuttavia, la modellizzazione di questi sistemi presenta alcune sfide computazionali. Tra queste, una delle difficoltà principali consiste nel simulare con precisione la dinamica delle particelle all’interno del plasma, un compito che richiede modelli ibridi, a metà tra fluidodinamica e cinetica. A differenza dei tokamak, la geometria non assisimmetrica degli stellarator induce un accoppiamento intrinseco tra i modi toroidali, che complicano ulteriormente la simulazione.

Partendo da tali premesse, il presente lavoro propone, dunque, un nuovo schema di proiezione, integrato nel codice a elementi finiti JOREK. Tale approccio consente di trasferire l'informazione cinetica sulla mesh in modo consistente e fisicamente fondato, gestendo la complessità della mesh 3D. Ciò avviene costruendo e risolvendo un unico sistema lineare che include simultaneamente tutte le armoniche toroidali. 

Al fine di gestire in modo efficiente e gestibile la localizzazione di milioni di particelle, è stato sviluppato un algoritmo di indirizzamento spaziale attraverso un R-Tree tridimensionale, il quale garantisce la fattibilità computazionale del metodo su larga scala. L’affidabilità del codice sviluppato è stata rigorosamente verificata su geometrie realistiche del reattore stellarator Wendelstein 7-X. 

I test di convergenza hanno confermato il raggiungimento della convergenza spettrale prevista dalla teoria, evidenziando, inoltre, il contrasto con le scarse prestazioni degli approcci non accoppiati. Di conseguenza, si è evidenziato come lo sviluppo di questo strumento computazionale fornisca una risorsa essenziale per l'analisi predittiva e l'ottimizzazione di futuri progetti cinetici su stellarator.

\vspace{15pt}

\begin{tcolorbox}[arc=0pt, boxrule=0pt, colback=bluePoli!60, width=\textwidth, colupper=white]
    \textbf{Parole chiave:} Simulazioni di stellarator, Modelli ibridi fluido-cinetici, Metodo degli Elementi Finiti (FEM), Metodo Particle-in-Cell (PiC), JOREK, Calcolo ad Alte Prestazioni (HPC) 
\end{tcolorbox}

\end{document}